\documentclass[aps,pra,superscriptaddress,reprint,floatfix,longbibliography,showpacs,amsfonts,amsmath,amssymb]{revtex4-1}
\usepackage{graphicx}
\usepackage{hyperref}
\usepackage{xr-hyper}
\usepackage{upgreek}
\usepackage{epstopdf}
\usepackage[note-name=, use-sort-key = false]{notes2bib}
\usepackage{color}

\begin{document}

\title{Signatures of van Hove singularities in the anisotropic in-plane optical conductivity of the topological semimetal Nb$_3$SiTe$_6$}

\author{J. Ebad-Allah}
\affiliation{Experimentalphysik II, Institute for Physics, University of Augsburg, D-86135 Augsburg, Germany}
\affiliation{Department of Physics, Tanta University, 31527 Tanta, Egypt}
\author{A. A. Tsirlin}
\affiliation{Felix Bloch Institute for Solid-State Physics, Leipzig University, 04103 Leipzig, Germany}
\affiliation{Experimentalphysik VI, Center for Electronic Correlations and Magnetism,
Institute for Physics, University of Augsburg, D-86135 Augsburg, Germany}
\author{Y. L. Zhu}
\affiliation{2D Crystal Consortium, Materials Research Institute, Pennsylvania State University, University Park, PA 16802, USA}
\affiliation{Department of Physics, Pennsylvania State University, University Park, Pennsylvania 16802, USA}
\author{Z. Q. Mao}
\affiliation{2D Crystal Consortium, Materials Research Institute, Pennsylvania State University, University Park, PA 16802, USA}
\affiliation{Department of Physics, Pennsylvania State University, University Park, Pennsylvania 16802, USA}
\author{C. A. Kuntscher}
\email{christine.kuntscher@physik.uni-augsburg.de}
\affiliation{Experimentalphysik II, Institute for Physics, University of Augsburg, D-86135 Augsburg, Germany}

\begin{abstract}
We present a temperature-dependent infrared spectroscopy study on the layered topological semimetal Nb$_3$SiTe$_6$ combined with density-functional theory (DFT) calculations of the electronic band structure and optical conductivity. Our results reveal an anisotropic behavior of the in-plane ($ac$-plane) optical conductivity, with three pronounced excitations located at around 0.15, 0.28, and 0.41~eV for the polarization of the incident radiation along the $c$ axis. These excitations are well reproduced in the theoretical spectra. Based on the \textit{ab initio} results, the excitations around 0.15 eV and 0.28 eV are interpreted as fingerprints of van Hove singularities in the electronic band structure and compared to the findings for other topological semimetals.
\end{abstract}


\maketitle

\section{Introduction}

\begin{figure}[t]
\includegraphics[width=0.35\textwidth]{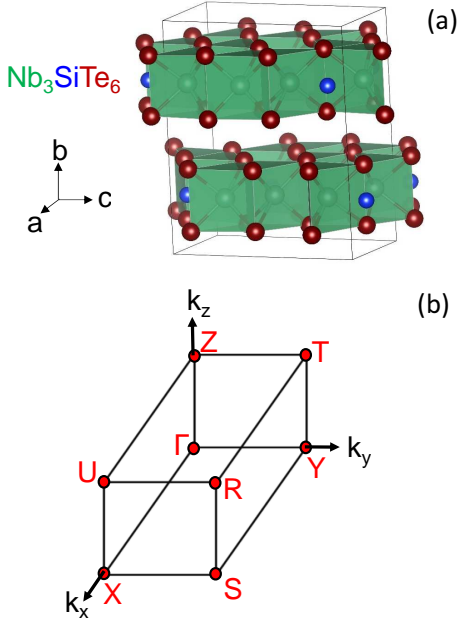}
\caption{(a) Crystal structure of  Nb$_3$SiTe$_6$. (b) First Brillouin zone of Nb$_3$SiTe$_6$ with the notation of the high-symmetry points.}
\label{fig:crystalstructure}
\end{figure}

Novel layered Dirac materials hosting nontrivial band crossings in the vicinity of the Fermi level (E$_{F}$) attract considerable attention in condensed-matter research due to their unusual physical properties and phenomena such as anisotropic electron transport \cite{Hu.2015}, chiral anomaly \cite{Son.2013}, nodal chains \cite{Wang.2017}, hourglass dispersions \cite{Li.2018}, drumhead-like states \cite{Yu.2017}, van Hove singularities \cite{Chen.2015,Biswas.2020,Heumen.2019,Kim.2018,Neupane.2015,Havener.2014},
and surface superconductivity \cite{Heikkilae.2011}.
The layered ternary telluride compounds $\textit{M}_3$SiTe$_6$ ($\textit{M}$= Nb and Ta) are one class of these materials, where the band structure calculations predicted several nontrivial band features  near E$_{F}$ \cite{Li.2018,Sato.2018,Naveed.2020}.
Nb$_3$SiTe$_6$ is a van-der-Waals layered material with the crystal structure very similar to that of MoS$_{2}$ \cite{Lee.2010}, and can be thinned down to atomically thin 2D crystals \cite{Hu.2015,Zhu.2020}. Bulk Nb$_3$SiTe$_6$ has an orthorhombic symmetry with the space group $Pnma$ \cite{Ohno.1999}. The layers stack via van-der-Waals forces, forming bundles of sandwich layers with the order Te-(Nb,Si)-Te.
Each Te-(Nb,Si)-Te layer is composed of face- and edge-sharing NbTe$_6$ prisms with Si ions inserted into interstitial sites among these prisms, as illustrated in Fig.\ \ref{fig:crystalstructure}(a). We also depict in Fig.\ \ref{fig:crystalstructure}(b) the first Brillouin zone of bulk  Nb$_3$SiTe$_6$ with the high-symmetry points.

In the absence of the spin-orbit coupling (SOC), the electronic band structure of $\textit{M}_3$SiTe$_6$ contains (i) a nodal loop related to the linear-band-crossing points along the $\Gamma-Y$ and $\Gamma-Z$ paths,  and (ii) a fourfold nodal line formed along the $S-R$ path \cite{Li.2018,Sato.2018,Naveed.2020}. Adding SOC leads to several new features, for instance: (i) gapping the nodal loop around the $\Gamma$ point, (ii) fourfold degeneracy of each band along the paths $U-X$, $R-U$, and $Z-S$, and (iii) the emergence of an hourglass Dirac loop along the $S-R$ path instead of the nodal line as well as along $S-X$. The degeneracies of features (ii) and (iii) result from the nonsymmorphic space group symmetry. Furthermore, the charge carrier mobility along the $\textit{a}$ direction was predicted to be much higher due to the Dirac dispersion along this direction, leading to a strong anisotropy in the electronic properties.

Temperature-dependent resistivity measurements on a bulk sample of Nb$_3$SiTe$_6$ showed typical metallic behavior with an anisotropy along in-plane and out-of-plane directions, related to the specific bonding state of Nb ions. In particular, according to the projected band structure and density of states the conduction bands crossing the Fermi level are mainly derived from Nb $4d$ orbitals \cite{Hu.2015}. Furthermore, an in-plane anisotropy is also expected for the $\textit{M}_3$SiTe$_6$ compounds due to the large difference between the in-plane lattice parameters \cite{Ohno.1999}. Consistently, an angle-resolved photoemission spectroscopy study reported a strong anisotropy of the Fermi surface of Ta$_3$SiTe$_6$ \cite{Sato.2018}.
In Nb$_3$SiTe$_6$, both hole and electron pockets exist, hence in the bulk the presence of free charge carriers with different scattering rates is expected. The hole-type charge carriers are suggested to prevail in the transport properties of thin flakes \cite{An.2018}, while their density will be affected by the deficiency of Te atoms \cite{Pang.2020}.

Despite Nb$_3$SiTe$_6$ revealing very interesting electronic properties, its optical conductivity has not been studied yet. In this paper we investigate the temperature-dependent in-plane ($ac$-plane) optical conductivity of bulk Nb$_3$SiTe$_6$ single crystal, obtained by frequency-dependent reflectivity measurements for the polarization directions {\bf E}$\| a$ and {\bf E}$\| c$. The optical conductivity shows an anisotropic behaviour between the two in-plane polarization directions. Several pronounced interband excitations are observed in the optical conductivity along the $c$ axis, which sharpen as the temperature decreases. Based on density-functional theory (DFT) calculations
we relate these excitation peaks to specific transitions between electronic bands and propose that they are related to van Hove singularities in the electronic band structure.

\begin{figure*}
\includegraphics[width=0.85\textwidth]{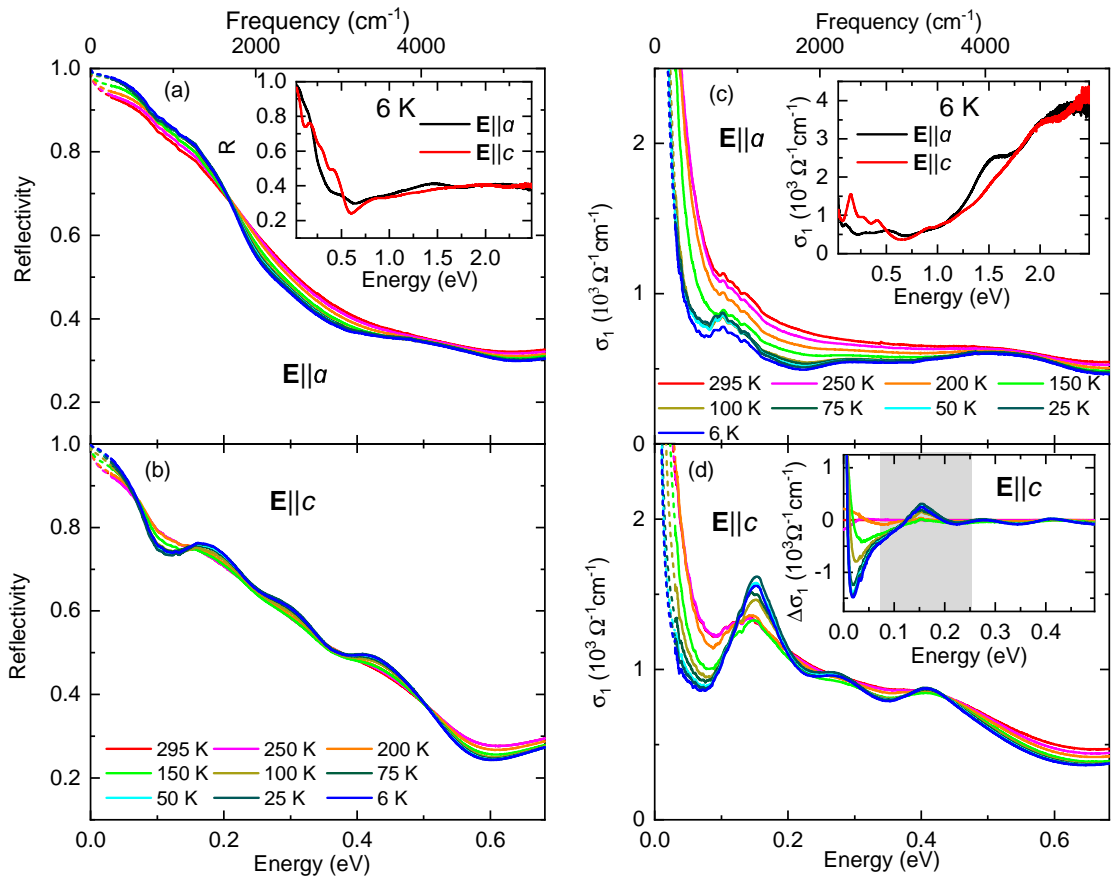}
\caption{Temperature-dependent reflectivity spectra of Nb$_3$SiTe$_6$ for the polarization directions (a) {\bf E}$\| a$ and (b) {\bf E}$\| c$. Inset of (a): Comparison between the reflectivity spectra at 6 K for the two polarization directions over a broad frequency range. (c) and (d) Temperature-dependent optical conductivity $\sigma_1$ spectra of Nb$_3$SiTe$_6$ for {\bf E}$\| a$ and {\bf E}$\| c$, respectively. Inset of (c): Comparison between the $\sigma_1$ spectra at 6 K for the two polarization directions in the entire measured range. Inset of (d): Difference spectra $\Delta\sigma_1$ calculated according to $\Delta\sigma_{1}(\omega, T)= \sigma_{1}(\omega, T)-\sigma_{1}(\omega, 295~K)$. The gray area marks the full-width-at-half-maximum of the L1 peak at 295~K.}
\label{fig:opticalconductivity}
\end{figure*}

\section{SAMPLE PREPARATION AND EXPERIMENTAL DETAILS}

Single crystals of Nb$_{3}$SiTe$_{6}$ were grown using chemical vapor transport with a mixture of Nb, Si, ant Te at a molar ratio of 3:1:6. During synthesis, the temperature of the hot and cold ends of the double-zone tube furnace was set at 950$^{\circ}$C and 850$^{\circ}$C, respectively \cite{Hu.2015, An.2018}.

The temperature-dependent reflectivity measurements at ambient pressure were performed between 295 and 6~K in the frequency range from 0.025 to 2.48~eV (200 to 20000~cm$^{-1}$). Measurements have been conducted on a single crystal on a freshly cleaved {\bf E}$\| ac$ surface. A silver
layer was evaporated onto half of the sample surface to serve as a reference, and the obtained spectra have been corrected with the mirror reflectivity later on. The sample was mounted on a cold-finger microcryostat. The measurements were carried out for the in-plane polarization directions {\bf E}$\| a$ and {\bf E}$\| c$ using an infrared microscope (Bruker Hyperion), equipped with a 15$\times$ Cassegrain objective, coupled to a Bruker Vertex 80v FT-IR spectrometer.

The Kramers-Kronig (KK) relations were applied to transform the reflectivity spectra into the complex optical conductivity $\sigma (\omega)=\sigma_{1} (\omega)+i \sigma_{2} (\omega)$  and the complex dielectric function $\epsilon (\omega)=\epsilon_{1} (\omega)+ i \epsilon_{2} (\omega)$. The extrapolation of the reflectivity data were done in a manner similar to our previous publications \cite{EbadAllah.2019,Koepf.2020,EbadAllah.2021}. To this end, the reflectivity was extrapolated to low frequencies based on a Drude-Lorentz fit, while for the high-frequency extrapolation we used the x-ray atomic scattering functions \cite{Tanner.2015}.
To obtain the contributions to the optical conductivity, the reflectivity and optical conductivity spectra were simultaneously fitted with the Drude-Lorentz model.

Band structure of Nb$_3$SiTe$_6$ was calculated in the Wien2K code~\cite{wien2k,Blaha2020} using the Perdew-Burke-Ernzerhof (PBE) type of the exchange-correlation potential~\cite{pbe96}. Lattice parameters and atomic positions from Ref.~\cite{Li.1992} were employed without further optimization. The corresponding notation of the high-symmetry points is shown in Fig.~\ref{fig:crystalstructure}(b). Charge density was converged on the $8\times 4\times 4$ $k$-mesh. Consequently, optical conductivity was calculated with the internal routines of Wien2K~\cite{Draxl2006} on the dense $24\times 12\times 12$ mesh.

\section{RESULTS AND DISCUSSION}

The temperature-dependent reflectivity spectra of Nb$_3$SiTe$_6$ for the in-plane polarization directions {\bf E}$\| a$ and {\bf E}$\| c$ are shown in Fig.\ \ref{fig:opticalconductivity} (a) and (b), respectively, and in Fig.\ S1 in the Supplemental Material \cite{Suppl}. The plasma edge in the reflectivity and the increase of the low-frequency reflectivity level during cooling down, which almost approaches unity at 6~K for both polarization directions, reveals the metallic nature of the compound consistent with transport measurements \cite{Hu.2015,Sato.2018,An.2018,Pang.2020}.
The plasma edge in the reflectivity depends not only on the temperature but also on the polarization direction as illustrated in the inset of Fig\ \ref{fig:opticalconductivity} (a).
Figures\ \ref{fig:opticalconductivity} (c) and (d)  display the temperature-dependent real part of the optical conductivity $\sigma_1$ for both polarization directions, as derived from the reflectivity spectra through KK relations. Corresponding plots on a lin-log scale can be found in Fig.\ S1 in the Supplemental Material \cite{Suppl}.  For both in-plane polarization directions, the $\sigma_1$ spectrum shows intraband excitations at low frequencies described by Drude contributions, which become sharper during cooling down due to reduced scattering.

For the further analysis and discussion of other excitations (besides the intraband transitions) we divide the measured energy range into two regions: (i) the low-energy region (energies between 0.08 eV and 0.7~eV) and (ii) the high-energy region (energies from 0.7 eV up to 2.25 eV).
In the low-energy region, the $\sigma_1$ spectrum along both axes shows a drop at around 0.1 eV followed by several polarization-dependent interband excitations at energies below 0.7 eV.  Interestingly, most of the observed excitations in this energy range hardly shift with decreasing temperature but only sharpen.
This temperature evolution of the low-energy excitations can be explained by a simple approach taking the temperature dependence of the Fermi-Dirac distribution function into account, causing a sharpening of the low-energy interband transitions with cooling. This approach was recently demonstrated for the Weyl semimetal TaP \cite{Yaresko.2021}.
In the high-energy region, $\sigma_1$ exhibits a monotonic increase with increasing frequency overlaid with various excitations, whose positions depend on the polarization direction. The overall changes in the optical conductivity in this energy region during cooling down are modest for both $a$ and $c$ axis.

\begin{figure} [t]
\centering
\includegraphics[width=0.4\textwidth]{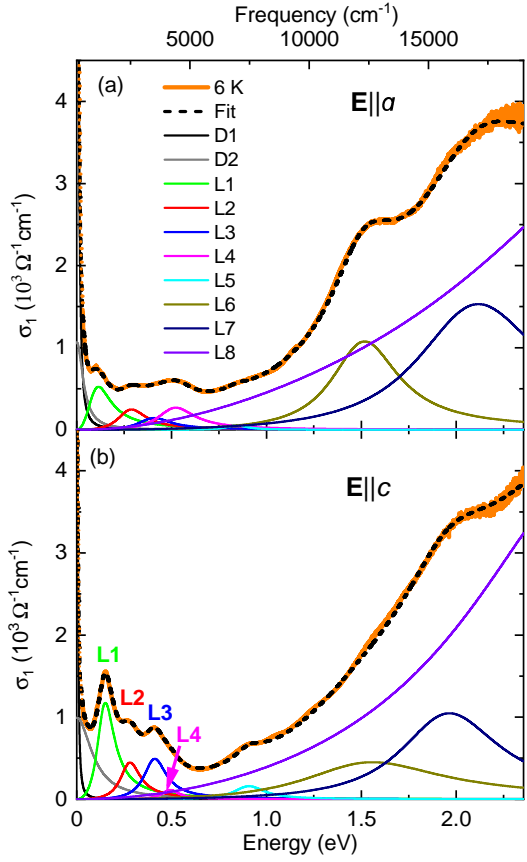}
\caption{Real part of the optical conductivity $\sigma_1$ of Nb$_3$SiTe$_6$ at 6 K for (a) {\bf E}$\| a$ and (b) {\bf E}$\| c$ together with the total Drude-Lorentz fit and the various fitting contributions (D: Drude term, L: Lorentz term).}
		\label{fig:sigma-fit}
\end{figure}

\begin{figure} [t]
\centering
\includegraphics[width=0.5\textwidth]{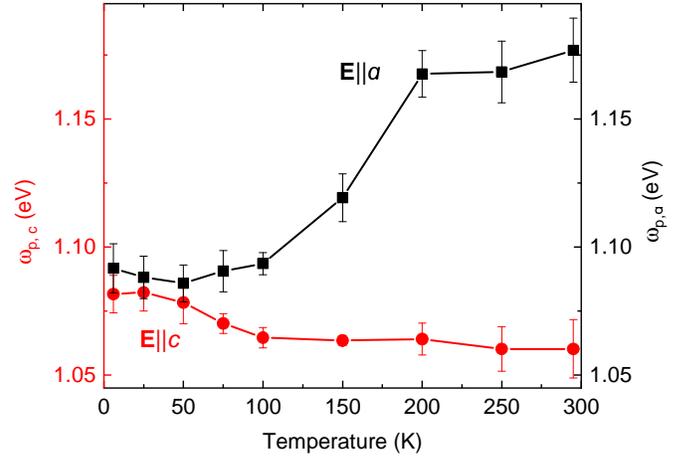}
\caption{Temperature-dependent effective plasma frequency of Nb$_3$SiTe$_6$ for the polarization directions {\bf E}$\| a$ ($\omega_{p,a}$) and  {\bf E}$\| c$ ($\omega_{p,c}$).}
		\label{fig:plasmafrequency}
	\end{figure}

\begin{figure*}
\includegraphics[width=1\textwidth]{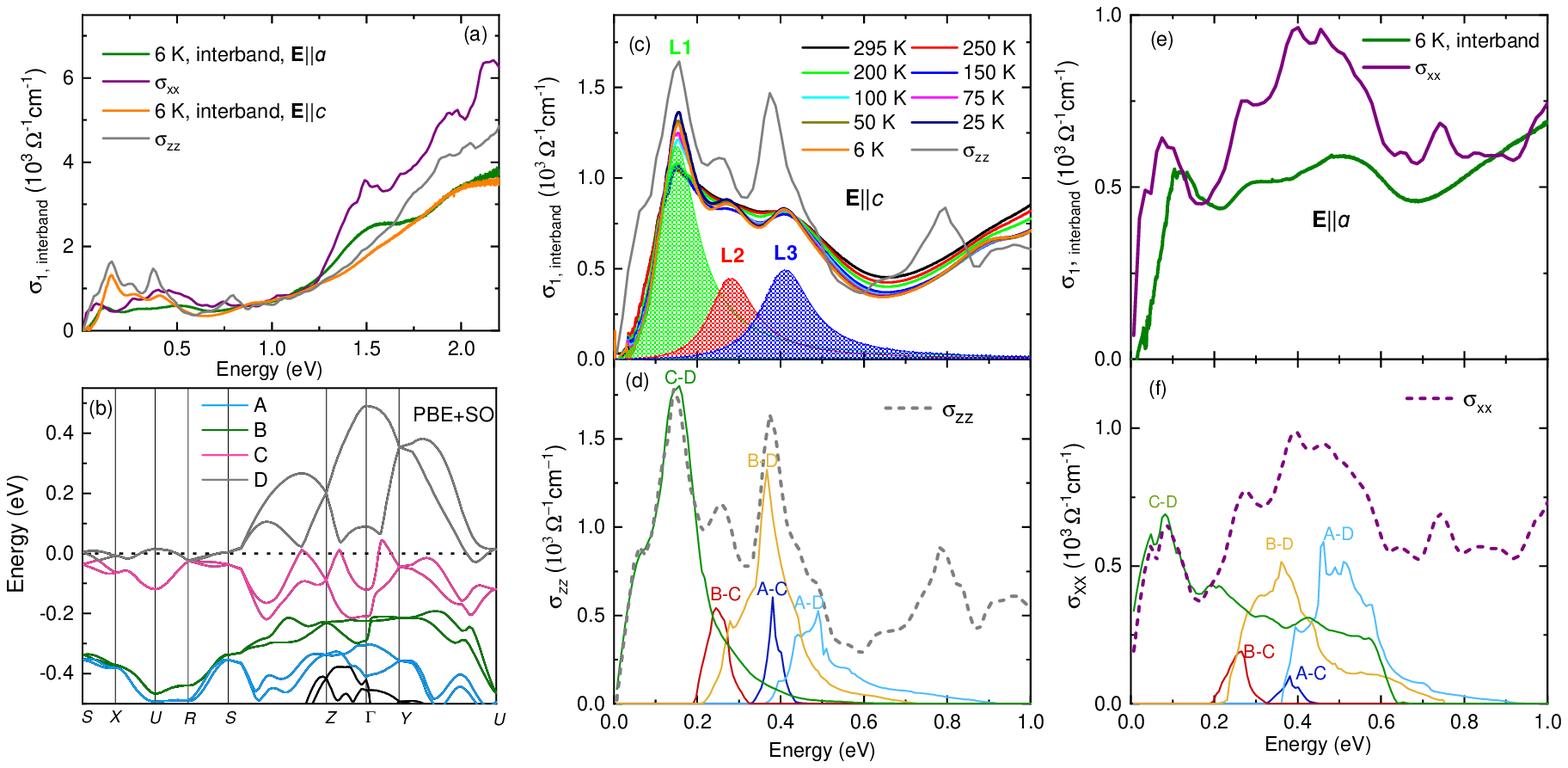}
\caption{(a) Comparison between the experimental and DFT interband conductivity $\sigma_{1, interband}$ for energies up to 2.2 eV for the polarization directions {\bf E}$\| a$ and {\bf E}$\| c$ at 6 K. (b) Calculated band structure of Nb$_3$SiTe$_6$ with SOC.
(c) Temperature-dependent experimental interband conductivity $\sigma_{1, interband}$ below 1.0 eV for the polarization directions {\bf E}$\| c$ compared with the DFT interband conductivity $\sigma_{zz}$. The colored peaks are the fitting contributions for L1, L2, L3 using the Lorentz model.
(d) Contributions of different band combinations to the optical conductivity $\sigma_{zz}$.
(e) The experimental $\sigma_{1, interband}$ at 6 K for the polarization
{\bf E}$\| a$ compared with the DFT interband conductivity $\sigma_{xx}$.
(f) Contributions of different band combinations to the optical conductivity $\sigma_{xx}$.}
\label{fig:theory}
\end{figure*}

Focusing on the experimental spectra at the lowest studied temperature 6~K, one observes that the optical response contains several pronounced contributions for {\bf E}$\| c$ compared to {\bf E}$\| a$, as illustrated in the insets of Figs.\ \ref{fig:opticalconductivity}(a) and (c). In order to distinguish the optical contributions of each axis, we fitted the $\sigma_{1}$ spectra (and simultaneously the reflectivity spectra) using the Drude-Lorentz model. As an example, we display in the two panels of Fig.\ \ref{fig:sigma-fit} the fitting of the $\sigma_{1}$ spectrum together with the fitting contributions for {\bf E}$\| a$ and {\bf E}$\| c$ at 6 K (see also Figs.\ S2 and S3 in the Supplemental Material \cite{Suppl}). From the fitting we obtained the following contributions for the polarization {\bf E}$\| c$: (i) Below 0.7 eV, the $\sigma_1$ spectrum consists of two Drude contributions related to the itinerant charge carriers \cite{Pang.2020}.
The presence of different types of free carriers is reasonable according to the reported electronic band structure, as already mentioned in the Introduction. In fact, two Drude components had to be included in the model for the {\bf E}$\| c$ polarization direction to obtain a reasonable fit quality.  Most interestingly, the E$\| c$ $\sigma_1$ spectrum shows a sharp peak-like excitation (L1) at around 0.15 eV, followed by two less sharp but still pronounced excitations at around 0.28 and 0.41~eV (L2 and L3, respectively), and a shoulder (L4) at around 0.49 eV. The main interband contributions (L1 -- L3) to the {\bf E}$\| c$ optical conductivity are also highlighted in Fig.\ \ref{fig:theory}(c).
(ii) Above 0.7~eV, another three high-energy excitations overlay the monotonic increase, and are positioned at approximately  0.93 eV, 1.5 eV, and 2.0 eV [see Fig.\ \ref{fig:sigma-fit}(b)].

For the polarization direction {\bf E}$\| a$ [see Fig.\ \ref{fig:sigma-fit}(a)], one Drude term was sufficient for obtaining a good fit quality.  However, for consistency reasons, we included two Drude contributions for this polariziation direction as well, like for the {\bf E}$\| c$ optical spectrum, accounting for different types of carriers.
The observed interband excitations below 0.7 eV are smeared out compared to {\bf E}$\| c$. In addition, most of the observed excitations below and above 0.7 eV are located at slightly different energies, namely at around 0.12 eV, 0.29 eV, 0.41 eV, and 0.52 eV in the low-energy region, and at around 0.9 eV, 1.5 eV, and 2.2 eV in the high-energy region [see Fig.\ \ref{fig:sigma-fit}(a) for the fitting contributions]. These differences between the observed excitations for the two directions give a clear evidence for the in-plane anisotropy of Nb$_3$SiTe$_6$, consistent with the in-plane anisotropy of the crystal structure, which is further illustrated by the large difference between the in-plane lattice parameters $a$ and $c$ ($a$=6.353 ${\AA}$, $b$=11.507 ${\AA}$, $c$=13.938 ${\AA}$ according to Ref.\ \cite{Li.1992}). An in-plane anisotropy was also observed in the Fermi surface of the sister compound Ta$_3$SiTe$_6$ \cite{Sato.2018}.

The in-plane anisotropy is also revealed by the effective plasma frequency $\omega_{p}$, which was calculated from the plasma frequencies $\omega_{p1}$ and $\omega_{p2}$ of the two Drude contributions according to $\omega_{p}=\sqrt{\omega_{p1}^{2}+\omega_{p2}^{2}}$.
The temperature-dependent plasma frequency for the two measured polarization directions is shown in Fig.\ \ref{fig:plasmafrequency}.
For ${\bf E}\| a$ $\omega_{p}$ decreases as the temperature decreases, namely from  $\omega_{p,300 K}$ = 1.18 eV to $\omega_{p,6 K}$ = 1.09 eV. Such a temperature dependence is expected for a semimetal, with less free carriers at low temperatures \cite{Crassee.2018}.
Along the $c$ axis, $\omega_{p}$ exhibits an unusual behaviour, where it slightly increases from $\omega_{p,300 K}$ = 1.05 eV to $\omega_{p,6 K}$ = 1.11 eV upon cooling (see Fig.\ \ref{fig:plasmafrequency}). Generally, a direct relation should be expected between the electronic kinetic energy and the Drude spectral weight, which is directly proportional to $\omega_{p}^{2}$. An increase in $\omega_{p}$ with cooling thus indicates an increase in electronic kinetic energy, which can be due to reduced electron correlation effects, i.e., reduced effective electronic mass \cite{Xu.2020, Shao.2020}. Another possible reason could be a temperature-induced shift of the Fermi level, as was recently demonstrated for the semimetal ZrTe$_5$ \cite{Xu.2018}. However, a Fermi level shift should affect the plasma frequency for both polarization directions in the same manner, which is inconsistent with our results.

The temperature-dependent $\sigma_1$ spectrum for {\bf E}$\| c$ suggests a considerable redistribution of the spectral weight from low to high frequencies upon cooling [see Fig.\ \ref{fig:opticalconductivity}(d)].
The difference spectra $\Delta\sigma_1$, defined as $\Delta\sigma_{1}(\omega, T)= \sigma_{1}(\omega, T)-\sigma_{1}(\omega, 295 K)$, illustrate the temperature-induced reshuffling of the spectral weight among various contributions. Such kind of analysis is in particular interesting for the ${\bf E}\| c$ optical conductivity spectrum with the most pronounced excitations. We observe that the spectral weight redistribution mainly occurs between the Drude contributions and the mid-infrared excitations [see inset of Fig.\ \ref{fig:opticalconductivity} (d)]. At first sight, it seems that some Drude spectral weight is transferred to the L1 excitation during cooling down. However, when taking into account the full-width-at-half-maximum of the L1 peak at room temperature, as indicated by the grey area, it is clear that the spectral weight of the L1 peak is decreased in certain energy ranges during cooling. For better illustration, we plot in Fig.\ S4 in the Supplemental Material \cite{Suppl} the difference spectra $\Delta\sigma_1$ together with the temperature-dependent L1 Lorentz peak. From our detailed fitting analysis of the experimental data we find that the Drude spectral weight is slightly increasing with decreasing temperature (increasing plasma frequency $\omega_{p}$ for {\bf E}$\| c$, see Fig.\ \ref{fig:plasmafrequency}), whereas the spectral weight of the L1 peak (as well as the L2 peak) is decreasing, as will be discussed in more detail later.
The transfer of the spectral weight indicates the reconstruction of the electronic energy bands over the energy range below 0.7~eV.

\begin{figure}[t]
\includegraphics[width=0.5\textwidth]{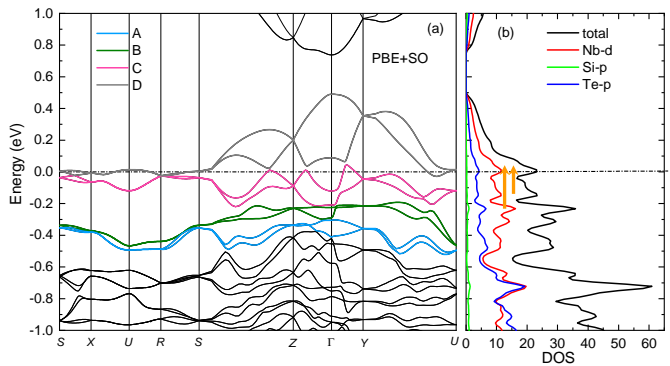}
\caption{(a) Calculated electronic band structure of Nb$_3$SiTe$_6$.
(b) Total density of states together with the partial density of states for Nb $4d$, Si $3p$, and Te $5p$. The two vertical arrows mark the possible electronic transitions, which could explain the two low-frequency absorption peaks L1 and L2 in the {\bf E}$\| c$ optical conductivity.}
\label{fig:DOS}
\end{figure}

For the interpretation of the observed excitations in the experimental optical conductivity spectra, we carried out DFT electronic band structure calculations and calculations of the optical conductivity. Hereby, SOC was taken into account.
The calculated electronic band structure, as depicted in Fig.\ \ref{fig:theory}(b), is in agreement with earlier results \cite{Li.2018}. Namely, we observe the gapping of the nodal loop around $\Gamma$ point, the appearance of an hourglass Dirac loop along the $S-R$ and $S-X$ paths, and the degeneracy of the bands along several paths.
The electronic band structure contains several nontrivial bands near E$_F$, mainly originating from  Nb $4d$ orbitals \cite{Hu.2015}, as evidenced by the partial density of states depicted in Fig.\ \ref{fig:DOS} (b).
For the further discussion, we group adjacent electronic bands into pairs, labeled A, B, C, and D, with the C and D bands lying in direct vicinity of $E_{F}$.
The theoretical plasma frequencies from the DFT calculations amount to 0.138~eV for {\bf E}$\| a$, 0.699~eV for {\bf E}$\| b$, and 0.943~eV for {\bf E}$\| c$.
Accordingly, the one for {\bf E}$\| c$ matches the experimental value, while the one for {\bf E}$\| a$ is lower compared to the experimental result.

Figure \ref{fig:theory} (a) displays a comparison between the calculated interband conductivity $\sigma_{xx}$ and $\sigma_{zz}$ together with the corresponding experimental results $\sigma_{1, interband}$ at 6~K for the polarization directions {\bf E}$\| a$ and {\bf E}$\| c$, respectively, over a broad energy range. As DFT provides only the interband contribution, we subtracted the Drude terms from the experimental $\sigma_{1}$ spectra and show $\sigma_{1, interband}$. Obviously, the overall experimental interband conductivity spectra for both polarization directions are well reproduced by theory, especially in the low-energy region below 1\,eV, which will be the main focus in the following. The somewhat less favorable agreement above 1\,eV is probably caused by inaccuracies in the description of the excited states within DFT.
It is important to note that the agreement between theory and experiment is in particular obvious for the {\bf E}$\| c$ polarization direction, since the theoretical $\sigma_{zz}$ spectrum also shows a low-energy three-peak profile with peak energies 0.15, 0.26, and 0.38~eV, corresponding to the L1, L2, and L3 peaks at positions 0.15, 0.28, and 0.41~eV in the measured $\sigma_1$ spectrum.

The calculated optical conductivity reveals a tiny peak below 0.1~eV for both polarization directions, which is absent in the experimental spectra and most likely hidden behind the Drude contribution. This tiny peak might be due to transitions within the hourglass Dirac loop very close to E$_F$.
One furthermore observes that the experimental {\bf E}$\| c$ $\sigma_{1, interband}$ below the first sharp L1 excitation is approximately linear in frequency and temperature independent, although the L1 peak itself shows a significant temperature dependence in its intensity.
A similar behavior was recently observed for the nodal-line semimetal BaNiS$_2$ \cite{Santos-Cottin.2021}. Here, temperature-dependent peaks in the theoretical optical conductivity spectra, located at the high-energy limit of a temperature-independent isosbestic linear-in-frequency conductivity, were associated with van Hove singularities (VHS's). These VHS's are related to saddle points of the electronic bands which result from the connections between Dirac cones in the reciprocal space.
Although the sharp peaks were not fully developed in the experimental optical data of BaNiS$_2$, it was suggested that such sharp VHS's in $\sigma_1$ are signatures of open Dirac nodal lines \cite{Santos-Cottin.2021}. In case of BaNiS$_2$ a pronounced transfer of the spectral weight from the Drude contributions to the sharp peaks via the temperature-independent isosbestic line was observed during cooling down, in contrast to our findings.
It is important to note that the appearance of a VHS is not a generic feature of open Dirac nodal lines.
For example, for dispersive open Dirac nodal lines the linear-in-frequency conductivity may end with a flat spectral response without any noticeable peaks \cite{Ahn.2017,Shao2019}.

The presence of Dirac nodal lines or loops with corresponding two-dimensional Dirac electrons is expected to cause characteristic fingerprints in the optical response, namely a frequency-independent interband optical conductivity related to the transitions within the Dirac cones \cite{Carbotte2017,Mukherjee.2017,Ahn.2017}. It was furthermore shown that energy-dispersive nodal lines (in contrast to flat nodal lines) can cause a linear-in-frequency behavior in the optical conductivity, similar to semimetals with separate Dirac nodal points \cite{Shao2019}. As mentioned in the Introduction, the electronic band structure of Nb$_3$SiTe$_6$ contains several nodal lines and loops in the vicinity of E$_F$. However, these nontrivial features are neither revealed in the experimental nor in the theoretical interband conductivity spectra [see Fig.\ \ref{fig:theory}(a)].

Based on the DFT calculations we can decompose the interband optical conductivity into contributions of different band combinations. In Figs.\ \ref{fig:theory}(d) and (f) we display the contributions of the interband transitions C-D, B-C, B-D, A-C, and A-D to the theoretical  conductivity spectra $\sigma_{zz}$ and $\sigma_{xx}$, respectively. In the following, we will focus on understanding the origin of these excitations and relate them to the observed excitation features in $\sigma_{1, interband}$.

By using the Drude-Lorentz fit, we managed to separate each contribution to the experimental $\sigma_{1}$ spectrum and to follow the respective temperature dependence.
A direct comparison between the fitting components L1, L2, L3, and L4 in the experimental $\sigma_{1, interband}$ spectra and the C-D, B-C, B-D, A-C and A-D excitations in the theoretical spectra reveals the following: (i) For {\bf E}$\| c$, the sharpest peak L1 originates from transitions between the C-D bands [Figs.\ \ref{fig:theory} (c) and (d)]. The L2 peak is due to transitions between the B-C bands, while the L3 peak results from transitions between A-C and B-D bands. The shoulder [Lorentz contribution L4, see Fig.\ \ref{fig:sigma-fit}(b)]
above the L3 peak can be associated with A-D transitions. However, based on the results of the DFT calculations we could not locate the positions in momentum space, where the transitions take place.
(ii) For {\bf E}$\| a$, the lowest-energy peak can be attributed to C-D transitions, like for {\bf E}$\| c$ [see Figs.\ \ref{fig:theory} (e) and (f)]. However, the interpretation of the higher-energy contributions to {\bf E}$\| a$ $\sigma_{1, interband}$ is less straightforward, since the C-D transitions contribute spectral weight over a rather broad frequency range, namely up to 0.6~eV, hence overlaying the contributions of other band combinations B-C, B-D etc. 

\begin{figure}[t]
\includegraphics[width=0.4\textwidth]{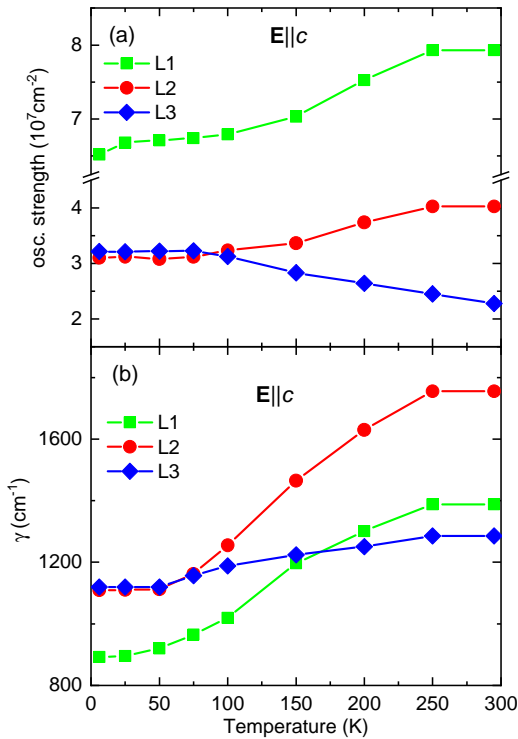}
\caption{Temperature-dependent (a) oscillator strength and (b) width ($\gamma$) of L1, L2, and L3 excitations for the polarization {\bf E}$\| c$, obtained from Drude-Lorentz fittings of $\sigma_1$.}
\label{fig:fitting}
\end{figure}

Further insight into the origin of these spectral features can be obtained from the electronic density of states (DOS) shown in Fig.~\ref{fig:DOS}(b). There are two DOS peaks located below the Fermi level and manifest VHS's. Another VHS is located right above E$_F$.
The accumulation of electronic states around these energies and the transitions between them [see arrows in Fig.~\ref{fig:DOS}(b)] could cause absorption peaks at energies 0.15 and 0.25~eV, which match the energy positions of the first two low-energy peaks in $\sigma_{zz}$, and hence could serve as an explanation for the L1 and L2 peaks in the experimental $\sigma_1$ spectrum [see Fig.\ \ref{fig:theory}(c)].
On the other hand, the fact that no DOS peak accompanies the higher-energy $\sigma_{zz}$ peaks (and thus L3 and L4) suggests a different origin of these features.

As we mentioned above, the L1, L2, and L3 excitations get sharper on decreasing temperature, while their energy positions are almost frequency-independent. Thus, by comparing the temperature dependence of the scattering rate (width) and oscillator strength of these three excitations we can gain insight whether these excitations have the same origin or not.
In Figs.\ \ref{fig:fitting} (a) and (b) we display the temperature-dependent scattering rate and oscillator strength, as extracted from the Drude-Lorentz fittings.
The L3 excitation shows a different behaviour compared to L1 and L2: whereas its width is hardly affected (slight narrowing) during cooling down, its oscillator strength monotonically increases with decreasing temperature and saturates below 100~K. This suggests that the L3 excitation has a different origin than the L1 and L2 excitations.
Interestingly, both L1 and L2 excitations display a very similar temperature behaviour: below 250~K their width as well as the oscillator strength monotonically decrease down to 100 K. The reduction in the width can be related to the reduced occupation of the bands D that lie immediately above E$_F$ and become less populated when temperature decreases. Below 100~K, both the scattering rate and oscillator strength for the L1 and L2 excitations are approximately constant. Since the L1 and L2 excitations show a similar temperature dependence, we conclude that they have a similar origin, contrasted with L3. This analysis underpins our assignment of L1 and L2 to the VHS features of the band structure.

The VHS interpretation of the L1 and L2 excitations is further supported by the low-dimensional character of the material, which leads to an increasing probability for the existence of VHS's in the electronic band structure, like in the layered Dirac semimetal ZrTe$_5$ \cite{Chen.2015}.
Additionally, the observed transfer of the spectral weight upon cooling, combined with the temperature-independent behaviour of $\sigma_{1, interband}$ at energies below L1, also suggest that these excitations are due to VHS's related to the nodal lines. As already mentioned above, a recent optical conductivity study of the nodal-line semimetal BaNiS$_{2}$ \cite{Santos-Cottin.2021} showed spectral weight transfer from low to high energies via a temperature-independent isosbestic line, ending at a VHS, which was proposed to result from the connections between Dirac cones in the reciprocal space.
Pronounced features in the optical conductivity spectra related to VHS's have also been observed in other Dirac materials, such as ZrTe$_5$ \cite{Chen.2015,Martino.2019}, PbTe$_{2}$ \cite{Heumen.2019}, $T_d$-MoTe$_2$ \cite{Santos-Cottin.2020}, and TaAs \cite{Santos-Cottin.2022}, and in the kagome metals Fe$_{3}$Sn$_{2}$ \cite{Biswas.2020}, Co$_3$Sn$_2$S$_2$ \cite{Xu.2020}, and AV$_3$Sb$_5$ with A = K, Rb, Cs \cite{Uykur.2021,Uykur2022,Wenzel2022}. These VHS's may have different origin. In ZrTe$_5$, the quasi-divergent peak observed in the optical conductivity was interpreted in terms of a VHS in the joint density of states. In AV$_3$Sb$_5$, they are due to band saddle points, while in Fe$_{3}$Sn$_{2}$ and Co$_3$Sn$_2$S$_2$ the observed sharp absorption peaks in the low-energy optical conductivity were attributed to the flat bands. In the case of the Dirac semimetal PbTe$_{2}$ the decrease and the collapse of the scattering rate of the low-energy charge carriers (i.e., the reduction in phase space for scattering) was interpreted as an experimental evidence for a VHS close to the Fermi level \cite{Heumen.2019}.

\section{CONCLUSION}

In summary, we have performed a polarization-dependent in-plane infrared spectroscopy study of Nb$_3$SiTe$_6$ at low temperature, combined with DFT calculations of the electronic band structure and optical conductivity. The comparative experimental and theoretical study revealed a similar profile of the interband optical conductivity for both in-plane polarization directions, namely, several peaks due to interband transitions followed by a monotonic increase overlaid with various high-energy excitations. We found that the interband excitations along the $c$ axis lead to pronounced and sharp peaks, in contrast to the less pronounced excitations for {\bf E}$\| a$, indicating an in-plane anisotropic behavior of Nb$_3$SiTe$_6$. Based on calculations of the band structure and optical conductivity, we assign the pronounced peaks at around 0.15 and 0.28~eV in the {\bf E}$\| c$ optical conductivity to van Hove singularities in the electronic density of states.

\begin{acknowledgments}
C.A.K. acknowledges financial support by the Deutsche Forschungsgemeinschaft (DFG), Germany, through Grant
No. KU 1432/15-1. Z.Q.M. acknowledges financial support by the US Department of Energy under grant DE-SC0019068.
\end{acknowledgments}


\begin{thebibliography}{45}%
\makeatletter
\providecommand \@ifxundefined [1]{%
 \@ifx{#1\undefined}
}%
\providecommand \@ifnum [1]{%
 \ifnum #1\expandafter \@firstoftwo
 \else \expandafter \@secondoftwo
 \fi
}%
\providecommand \@ifx [1]{%
 \ifx #1\expandafter \@firstoftwo
 \else \expandafter \@secondoftwo
 \fi
}%
\providecommand \natexlab [1]{#1}%
\providecommand \enquote  [1]{``#1''}%
\providecommand \bibnamefont  [1]{#1}%
\providecommand \bibfnamefont [1]{#1}%
\providecommand \citenamefont [1]{#1}%
\providecommand \href@noop [0]{\@secondoftwo}%
\providecommand \href [0]{\begingroup \@sanitize@url \@href}%
\providecommand \@href[1]{\@@startlink{#1}\@@href}%
\providecommand \@@href[1]{\endgroup#1\@@endlink}%
\providecommand \@sanitize@url [0]{\catcode `\\12\catcode `\$12\catcode
  `\&12\catcode `\#12\catcode `\^12\catcode `\_12\catcode `\%12\relax}%
\providecommand \@@startlink[1]{}%
\providecommand \@@endlink[0]{}%
\providecommand \url  [0]{\begingroup\@sanitize@url \@url }%
\providecommand \@url [1]{\endgroup\@href {#1}{\urlprefix }}%
\providecommand \urlprefix  [0]{URL }%
\providecommand \Eprint [0]{\href }%
\providecommand \doibase [0]{http://dx.doi.org/}%
\providecommand \selectlanguage [0]{\@gobble}%
\providecommand \bibinfo  [0]{\@secondoftwo}%
\providecommand \bibfield  [0]{\@secondoftwo}%
\providecommand \translation [1]{[#1]}%
\providecommand \BibitemOpen [0]{}%
\providecommand \bibitemStop [0]{}%
\providecommand \bibitemNoStop [0]{.\EOS\space}%
\providecommand \EOS [0]{\spacefactor3000\relax}%
\providecommand \BibitemShut  [1]{\csname bibitem#1\endcsname}%
\let\auto@bib@innerbib\@empty
\bibitem [{\citenamefont {Hu}\ \emph {et~al.}(2015)\citenamefont {Hu},
  \citenamefont {Liu}, \citenamefont {Yue}, \citenamefont {Liu}, \citenamefont
  {Zhu}, \citenamefont {He}, \citenamefont {Wei}, \citenamefont {Mao},
  \citenamefont {Antipina}, \citenamefont {Popov}, \citenamefont {Sorokin},
  \citenamefont {Liu}, \citenamefont {Adams}, \citenamefont {Radmanesh},
  \citenamefont {Spinu}, \citenamefont {Ji},\ and\ \citenamefont
  {Natelson}}]{Hu.2015}%
  \BibitemOpen
  \bibfield  {author} {\bibinfo {author} {\bibfnamefont {J.}~\bibnamefont
  {Hu}}, \bibinfo {author} {\bibfnamefont {X.}~\bibnamefont {Liu}}, \bibinfo
  {author} {\bibfnamefont {C.~L.}\ \bibnamefont {Yue}}, \bibinfo {author}
  {\bibfnamefont {J.~Y.}\ \bibnamefont {Liu}}, \bibinfo {author} {\bibfnamefont
  {H.~W.}\ \bibnamefont {Zhu}}, \bibinfo {author} {\bibfnamefont {J.~B.}\
  \bibnamefont {He}}, \bibinfo {author} {\bibfnamefont {J.}~\bibnamefont
  {Wei}}, \bibinfo {author} {\bibfnamefont {Z.~Q.}\ \bibnamefont {Mao}},
  \bibinfo {author} {\bibfnamefont {L.~Yu.}\ \bibnamefont {Antipina}}, \bibinfo
  {author} {\bibfnamefont {Z.I.}\ \bibnamefont {Popov}}, \bibinfo {author}
  {\bibfnamefont {P.~B.}\ \bibnamefont {Sorokin}}, \bibinfo {author}
  {\bibfnamefont {T.~J.}\ \bibnamefont {Liu}}, \bibinfo {author} {\bibfnamefont
  {P.~W.}\ \bibnamefont {Adams}}, \bibinfo {author} {\bibfnamefont {S.~M.~A.}\
  \bibnamefont {Radmanesh}}, \bibinfo {author} {\bibfnamefont {L.}~\bibnamefont
  {Spinu}}, \bibinfo {author} {\bibfnamefont {H.}~\bibnamefont {Ji}}, \ and\
  \bibinfo {author} {\bibfnamefont {D.}~\bibnamefont {Natelson}},\ }\bibfield
  {title} {\enquote {\bibinfo {title} {Enhanced electron coherence in
  atomically thin {Nb$_3$SiTe$_6$}},}\ }\href@noop {} {\bibfield  {journal}
  {\bibinfo  {journal} {Nat. Phys.}\ }\textbf {\bibinfo {volume} {11}},\
  \bibinfo {pages} {471} (\bibinfo {year} {2015})}\BibitemShut {NoStop}%
\bibitem [{\citenamefont {Son}\ and\ \citenamefont {Spivak}(2013)}]{Son.2013}%
  \BibitemOpen
  \bibfield  {author} {\bibinfo {author} {\bibfnamefont {D.~T.}\ \bibnamefont
  {Son}}\ and\ \bibinfo {author} {\bibfnamefont {B.~Z.}\ \bibnamefont
  {Spivak}},\ }\bibfield  {title} {\enquote {\bibinfo {title} {{Chiral anomaly
  and classical negative magnetoresistance of Weyl metals}},}\ }\href@noop {}
  {\bibfield  {journal} {\bibinfo  {journal} {Phys. Rev. B}\ }\textbf {\bibinfo
  {volume} {88}},\ \bibinfo {pages} {104412} (\bibinfo {year}
  {2013})}\BibitemShut {NoStop}%
\bibitem [{\citenamefont {Wang}\ \emph {et~al.}(2017)\citenamefont {Wang},
  \citenamefont {Liu}, \citenamefont {Yu}, \citenamefont {Sheng},\ and\
  \citenamefont {Yang}}]{Wang.2017}%
  \BibitemOpen
  \bibfield  {author} {\bibinfo {author} {\bibfnamefont {Shan-Shan}\
  \bibnamefont {Wang}}, \bibinfo {author} {\bibfnamefont {Ying}\ \bibnamefont
  {Liu}}, \bibinfo {author} {\bibfnamefont {Zhi-Ming}\ \bibnamefont {Yu}},
  \bibinfo {author} {\bibfnamefont {Xian-Lei}\ \bibnamefont {Sheng}}, \ and\
  \bibinfo {author} {\bibfnamefont {Shengyuan~A.}\ \bibnamefont {Yang}},\
  }\bibfield  {title} {\enquote {\bibinfo {title} {{Hourglass Dirac chain metal
  in rhenium dioxide}},}\ }\href@noop {} {\bibfield  {journal} {\bibinfo
  {journal} {Nat. Commun.}\ }\textbf {\bibinfo {volume} {8}},\ \bibinfo {pages}
  {1844} (\bibinfo {year} {2017})}\BibitemShut {NoStop}%
\bibitem [{\citenamefont {Li}\ \emph {et~al.}(2018)\citenamefont {Li},
  \citenamefont {Liu}, \citenamefont {Wang}, \citenamefont {Yu}, \citenamefont
  {Guan}, \citenamefont {Sheng}, \citenamefont {Yao},\ and\ \citenamefont
  {Yang}}]{Li.2018}%
  \BibitemOpen
  \bibfield  {author} {\bibinfo {author} {\bibfnamefont {Si}~\bibnamefont
  {Li}}, \bibinfo {author} {\bibfnamefont {Ying}\ \bibnamefont {Liu}}, \bibinfo
  {author} {\bibfnamefont {Shan-Shan}\ \bibnamefont {Wang}}, \bibinfo {author}
  {\bibfnamefont {Zhi-Ming}\ \bibnamefont {Yu}}, \bibinfo {author}
  {\bibfnamefont {Shan}\ \bibnamefont {Guan}}, \bibinfo {author} {\bibfnamefont
  {Xian-Lei}\ \bibnamefont {Sheng}}, \bibinfo {author} {\bibfnamefont {Yugui}\
  \bibnamefont {Yao}}, \ and\ \bibinfo {author} {\bibfnamefont {Shengyuan~A.}\
  \bibnamefont {Yang}},\ }\bibfield  {title} {\enquote {\bibinfo {title}
  {{Nonsymmorphic-symmetry-protected hourglass Dirac loop, nodal line, and
  Dirac point in bulk and monolayer {$X$$_3$SiTe$_6$ ($X$ = Ta, Nb)}}},}\
  }\href@noop {} {\bibfield  {journal} {\bibinfo  {journal} {Phys. Rev. B}\
  }\textbf {\bibinfo {volume} {97}},\ \bibinfo {pages} {045131} (\bibinfo
  {year} {2018})}\BibitemShut {NoStop}%
\bibitem [{\citenamefont {Yu}\ \emph {et~al.}(2017)\citenamefont {Yu},
  \citenamefont {Fang}, \citenamefont {Dai},\ and\ \citenamefont
  {Weng}}]{Yu.2017}%
  \BibitemOpen
  \bibfield  {author} {\bibinfo {author} {\bibfnamefont {R.}~\bibnamefont
  {Yu}}, \bibinfo {author} {\bibfnamefont {Z.}~\bibnamefont {Fang}}, \bibinfo
  {author} {\bibfnamefont {X.}~\bibnamefont {Dai}}, \ and\ \bibinfo {author}
  {\bibfnamefont {H.}~\bibnamefont {Weng}},\ }\bibfield  {title} {\enquote
  {\bibinfo {title} {Topological nodal line semimetals predicted from
  first-principles calculations},}\ }\href@noop {} {\bibfield  {journal}
  {\bibinfo  {journal} {Front. Phys.}\ }\textbf {\bibinfo {volume} {12}},\
  \bibinfo {pages} {127202} (\bibinfo {year} {2017})}\BibitemShut {NoStop}%
\bibitem [{\citenamefont {Chen}\ \emph {et~al.}(2015)\citenamefont {Chen},
  \citenamefont {Zhang}, \citenamefont {Schneeloch}, \citenamefont {Zhang},
  \citenamefont {Li}, \citenamefont {Gu},\ and\ \citenamefont
  {Wang}}]{Chen.2015}%
  \BibitemOpen
  \bibfield  {author} {\bibinfo {author} {\bibfnamefont {R.~Y.}\ \bibnamefont
  {Chen}}, \bibinfo {author} {\bibfnamefont {S.~J.}\ \bibnamefont {Zhang}},
  \bibinfo {author} {\bibfnamefont {J.~A.}\ \bibnamefont {Schneeloch}},
  \bibinfo {author} {\bibfnamefont {C.}~\bibnamefont {Zhang}}, \bibinfo
  {author} {\bibfnamefont {Q.}~\bibnamefont {Li}}, \bibinfo {author}
  {\bibfnamefont {G.~D.}\ \bibnamefont {Gu}}, \ and\ \bibinfo {author}
  {\bibfnamefont {N.~L.}\ \bibnamefont {Wang}},\ }\bibfield  {title} {\enquote
  {\bibinfo {title} {{Optical spectroscopy study of the three-dimensional Dirac
  semimetal {ZrTe$_5$}}},}\ }\href@noop {} {\bibfield  {journal} {\bibinfo
  {journal} {Phys. Rev. B}\ }\textbf {\bibinfo {volume} {92}},\ \bibinfo
  {pages} {075107} (\bibinfo {year} {2015})}\BibitemShut {NoStop}%
\bibitem [{\citenamefont {Biswas}\ \emph {et~al.}(2020)\citenamefont {Biswas},
  \citenamefont {Iakutkina}, \citenamefont {Wang}, \citenamefont {Lei},
  \citenamefont {Dressel},\ and\ \citenamefont {Uykur}}]{Biswas.2020}%
  \BibitemOpen
  \bibfield  {author} {\bibinfo {author} {\bibfnamefont {A.}~\bibnamefont
  {Biswas}}, \bibinfo {author} {\bibfnamefont {O.}~\bibnamefont {Iakutkina}},
  \bibinfo {author} {\bibfnamefont {Q.}~\bibnamefont {Wang}}, \bibinfo {author}
  {\bibfnamefont {H.~C.}\ \bibnamefont {Lei}}, \bibinfo {author} {\bibfnamefont
  {M.}~\bibnamefont {Dressel}}, \ and\ \bibinfo {author} {\bibfnamefont
  {E.}~\bibnamefont {Uykur}},\ }\bibfield  {title} {\enquote {\bibinfo {title}
  {{Spin-Reorientation-Induced Band Gap in {Fe$_3$Sn$_2$} : Optical Signatures
  of Weyl Nodes}},}\ }\href@noop {} {\bibfield  {journal} {\bibinfo  {journal}
  {Phys. Rev. Lett.}\ }\textbf {\bibinfo {volume} {125}},\ \bibinfo {pages}
  {076403} (\bibinfo {year} {2020})}\BibitemShut {NoStop}%
\bibitem [{\citenamefont {van Heumen}\ \emph {et~al.}(2019)\citenamefont {van
  Heumen}, \citenamefont {Berben}, \citenamefont {Neubrand},\ and\
  \citenamefont {Huang}}]{Heumen.2019}%
  \BibitemOpen
  \bibfield  {author} {\bibinfo {author} {\bibfnamefont {Erik}\ \bibnamefont
  {van Heumen}}, \bibinfo {author} {\bibfnamefont {Maarten}\ \bibnamefont
  {Berben}}, \bibinfo {author} {\bibfnamefont {Linda}\ \bibnamefont
  {Neubrand}}, \ and\ \bibinfo {author} {\bibfnamefont {Yingkai}\ \bibnamefont
  {Huang}},\ }\bibfield  {title} {\enquote {\bibinfo {title} {{Scattering rate
  collapse driven by a van Hove singularity in the Dirac semimetal
  {PdTe$_2$}}},}\ }\href@noop {} {\bibfield  {journal} {\bibinfo  {journal}
  {Phys. Rev. Materials}\ }\textbf {\bibinfo {volume} {3}},\ \bibinfo {pages}
  {114202} (\bibinfo {year} {2019})}\BibitemShut {NoStop}%
\bibitem [{\citenamefont {Kim}\ \emph {et~al.}(2018)\citenamefont {Kim},
  \citenamefont {Kim}, \citenamefont {Kim}, \citenamefont {Kim}, \citenamefont
  {Park},\ and\ \citenamefont {Min}}]{Kim.2018}%
  \BibitemOpen
  \bibfield  {author} {\bibinfo {author} {\bibfnamefont {Kyoo}\ \bibnamefont
  {Kim}}, \bibinfo {author} {\bibfnamefont {Sooran}\ \bibnamefont {Kim}},
  \bibinfo {author} {\bibfnamefont {J.~S.}\ \bibnamefont {Kim}}, \bibinfo
  {author} {\bibfnamefont {Heejung}\ \bibnamefont {Kim}}, \bibinfo {author}
  {\bibfnamefont {J.-H.}\ \bibnamefont {Park}}, \ and\ \bibinfo {author}
  {\bibfnamefont {B.~I.}\ \bibnamefont {Min}},\ }\bibfield  {title} {\enquote
  {\bibinfo {title} {{Importance of the van Hove singularity in superconducting
  {PdTe$_2$}}},}\ }\href@noop {} {\bibfield  {journal} {\bibinfo  {journal}
  {Phys. Rev. B}\ }\textbf {\bibinfo {volume} {97}},\ \bibinfo {pages} {165102}
  (\bibinfo {year} {2018})}\BibitemShut {NoStop}%
\bibitem [{\citenamefont {Neupane}\ \emph {et~al.}(2015)\citenamefont
  {Neupane}, \citenamefont {Xu}, \citenamefont {Sankar}, \citenamefont
  {Gibson}, \citenamefont {Wang}, \citenamefont {Belopolski}, \citenamefont
  {Alidoust}, \citenamefont {Bian}, \citenamefont {Shibayev}, \citenamefont
  {Sanchez}, \citenamefont {Ohtsubo}, \citenamefont {Taleb-Ibrahimi},
  \citenamefont {Basak}, \citenamefont {Tsai}, \citenamefont {Lin},
  \citenamefont {Durakiewicz}, \citenamefont {Cava}, \citenamefont {Bansil},
  \citenamefont {Chou},\ and\ \citenamefont {Hasan}}]{Neupane.2015}%
  \BibitemOpen
  \bibfield  {author} {\bibinfo {author} {\bibfnamefont {Madhab}\ \bibnamefont
  {Neupane}}, \bibinfo {author} {\bibfnamefont {Su-Yang}\ \bibnamefont {Xu}},
  \bibinfo {author} {\bibfnamefont {R.}~\bibnamefont {Sankar}}, \bibinfo
  {author} {\bibfnamefont {Q.}~\bibnamefont {Gibson}}, \bibinfo {author}
  {\bibfnamefont {Y.~J.}\ \bibnamefont {Wang}}, \bibinfo {author}
  {\bibfnamefont {I.}~\bibnamefont {Belopolski}}, \bibinfo {author}
  {\bibfnamefont {N.}~\bibnamefont {Alidoust}}, \bibinfo {author}
  {\bibfnamefont {G.}~\bibnamefont {Bian}}, \bibinfo {author} {\bibfnamefont
  {P.~P.}\ \bibnamefont {Shibayev}}, \bibinfo {author} {\bibfnamefont {D.~S.}\
  \bibnamefont {Sanchez}}, \bibinfo {author} {\bibfnamefont {Y.}~\bibnamefont
  {Ohtsubo}}, \bibinfo {author} {\bibfnamefont {A.}~\bibnamefont
  {Taleb-Ibrahimi}}, \bibinfo {author} {\bibfnamefont {S.}~\bibnamefont
  {Basak}}, \bibinfo {author} {\bibfnamefont {W.-F.}\ \bibnamefont {Tsai}},
  \bibinfo {author} {\bibfnamefont {H.}~\bibnamefont {Lin}}, \bibinfo {author}
  {\bibfnamefont {Tomasz}\ \bibnamefont {Durakiewicz}}, \bibinfo {author}
  {\bibfnamefont {R.~J.}\ \bibnamefont {Cava}}, \bibinfo {author}
  {\bibfnamefont {A.}~\bibnamefont {Bansil}}, \bibinfo {author} {\bibfnamefont
  {F.~C.}\ \bibnamefont {Chou}}, \ and\ \bibinfo {author} {\bibfnamefont
  {M.~Z.}\ \bibnamefont {Hasan}},\ }\bibfield  {title} {\enquote {\bibinfo
  {title} {Topological phase diagram and saddle point singularity in a tunable
  topological crystalline insulator},}\ }\href@noop {} {\bibfield  {journal}
  {\bibinfo  {journal} {Phys. Rev. B}\ }\textbf {\bibinfo {volume} {92}},\
  \bibinfo {pages} {075131} (\bibinfo {year} {2015})}\BibitemShut {NoStop}%
\bibitem [{\citenamefont {Havener}\ \emph {et~al.}(2014)\citenamefont
  {Havener}, \citenamefont {Liang}, \citenamefont {Brown}, \citenamefont
  {Yang},\ and\ \citenamefont {Park}}]{Havener.2014}%
  \BibitemOpen
  \bibfield  {author} {\bibinfo {author} {\bibfnamefont {R.~W.}\ \bibnamefont
  {Havener}}, \bibinfo {author} {\bibfnamefont {Y.}~\bibnamefont {Liang}},
  \bibinfo {author} {\bibfnamefont {L.}~\bibnamefont {Brown}}, \bibinfo
  {author} {\bibfnamefont {L.}~\bibnamefont {Yang}}, \ and\ \bibinfo {author}
  {\bibfnamefont {J.}~\bibnamefont {Park}},\ }\bibfield  {title} {\enquote
  {\bibinfo {title} {{Van Hove Singularities and Excitonic Effects in the
  Optical Conductivity of Twisted Bilayer Graphene}},}\ }\href@noop {}
  {\bibfield  {journal} {\bibinfo  {journal} {Nano Lett.}\ }\textbf {\bibinfo
  {volume} {14}},\ \bibinfo {pages} {3353} (\bibinfo {year}
  {2014})}\BibitemShut {NoStop}%
\bibitem [{\citenamefont {Heikkil\"a}\ \emph {et~al.}(2011)\citenamefont
  {Heikkil\"a}, \citenamefont {Kopnin},\ and\ \citenamefont
  {Volovik}}]{Heikkilae.2011}%
  \BibitemOpen
  \bibfield  {author} {\bibinfo {author} {\bibfnamefont {T.~T.}\ \bibnamefont
  {Heikkil\"a}}, \bibinfo {author} {\bibfnamefont {N.~B.}\ \bibnamefont
  {Kopnin}}, \ and\ \bibinfo {author} {\bibfnamefont {G.~E.}\ \bibnamefont
  {Volovik}},\ }\bibfield  {title} {\enquote {\bibinfo {title} {Flat bands in
  topological media},}\ }\href@noop {} {\bibfield  {journal} {\bibinfo
  {journal} {JETP Lett.}\ }\textbf {\bibinfo {volume} {94}},\ \bibinfo {pages}
  {233} (\bibinfo {year} {2011})}\BibitemShut {NoStop}%
\bibitem [{\citenamefont {Sato}\ \emph {et~al.}(2018)\citenamefont {Sato},
  \citenamefont {Wang}, \citenamefont {Nakayama}, \citenamefont {Souma},
  \citenamefont {Takane}, \citenamefont {Nakata}, \citenamefont {Iwasawa},
  \citenamefont {Cacho}, \citenamefont {Kim}, \citenamefont {Takahashi},\ and\
  \citenamefont {Ando}}]{Sato.2018}%
  \BibitemOpen
  \bibfield  {author} {\bibinfo {author} {\bibfnamefont {Takafumi}\
  \bibnamefont {Sato}}, \bibinfo {author} {\bibfnamefont {Zhiwei}\ \bibnamefont
  {Wang}}, \bibinfo {author} {\bibfnamefont {Kosuke}\ \bibnamefont {Nakayama}},
  \bibinfo {author} {\bibfnamefont {Seigo}\ \bibnamefont {Souma}}, \bibinfo
  {author} {\bibfnamefont {Daichi}\ \bibnamefont {Takane}}, \bibinfo {author}
  {\bibfnamefont {Yuki}\ \bibnamefont {Nakata}}, \bibinfo {author}
  {\bibfnamefont {Hideaki}\ \bibnamefont {Iwasawa}}, \bibinfo {author}
  {\bibfnamefont {Cephise}\ \bibnamefont {Cacho}}, \bibinfo {author}
  {\bibfnamefont {Timur}\ \bibnamefont {Kim}}, \bibinfo {author} {\bibfnamefont
  {Takashi}\ \bibnamefont {Takahashi}}, \ and\ \bibinfo {author} {\bibfnamefont
  {Yoichi}\ \bibnamefont {Ando}},\ }\bibfield  {title} {\enquote {\bibinfo
  {title} {Observation of band crossings protected by nonsymmorphic symmetry in
  the layered ternary telluride {Ta$_3$SiTe$_6$}},}\ }\href@noop {} {\bibfield
  {journal} {\bibinfo  {journal} {Phys. Rev. B}\ }\textbf {\bibinfo {volume}
  {98}},\ \bibinfo {pages} {121111} (\bibinfo {year} {2018})}\BibitemShut
  {NoStop}%
\bibitem [{\citenamefont {Naveed}\ \emph {et~al.}(2020)\citenamefont {Naveed},
  \citenamefont {Fei}, \citenamefont {Bu}, \citenamefont {Bo}, \citenamefont
  {Shah}, \citenamefont {Chen}, \citenamefont {Zhang}, \citenamefont {Liu},
  \citenamefont {Wei}, \citenamefont {Zhang}, \citenamefont {Guo},
  \citenamefont {Xi}, \citenamefont {Rahman}, \citenamefont {Zhang},
  \citenamefont {Zhang}, \citenamefont {Wan},\ and\ \citenamefont
  {Song}}]{Naveed.2020}%
  \BibitemOpen
  \bibfield  {author} {\bibinfo {author} {\bibfnamefont {M.}~\bibnamefont
  {Naveed}}, \bibinfo {author} {\bibfnamefont {F.}~\bibnamefont {Fei}},
  \bibinfo {author} {\bibfnamefont {H.}~\bibnamefont {Bu}}, \bibinfo {author}
  {\bibfnamefont {X.}~\bibnamefont {Bo}}, \bibinfo {author} {\bibfnamefont
  {S.~A.}\ \bibnamefont {Shah}}, \bibinfo {author} {\bibfnamefont
  {B.}~\bibnamefont {Chen}}, \bibinfo {author} {\bibfnamefont {Y.}~\bibnamefont
  {Zhang}}, \bibinfo {author} {\bibfnamefont {Q.}~\bibnamefont {Liu}}, \bibinfo
  {author} {\bibfnamefont {B.}~\bibnamefont {Wei}}, \bibinfo {author}
  {\bibfnamefont {S.}~\bibnamefont {Zhang}}, \bibinfo {author} {\bibfnamefont
  {J.}~\bibnamefont {Guo}}, \bibinfo {author} {\bibfnamefont {C.}~\bibnamefont
  {Xi}}, \bibinfo {author} {\bibfnamefont {A.}~\bibnamefont {Rahman}}, \bibinfo
  {author} {\bibfnamefont {Z.}~\bibnamefont {Zhang}}, \bibinfo {author}
  {\bibfnamefont {M.}~\bibnamefont {Zhang}}, \bibinfo {author} {\bibfnamefont
  {X.}~\bibnamefont {Wan}}, \ and\ \bibinfo {author} {\bibfnamefont
  {F.}~\bibnamefont {Song}},\ }\bibfield  {title} {\enquote {\bibinfo {title}
  {{Magneto-transport and Shubnikov-de Haas oscillations in the layered ternary
  telluride topological semimetal candidate {Ta$_3$SiTe$_6$}}},}\ }\href@noop
  {} {\bibfield  {journal} {\bibinfo  {journal} {Appl. Phys. Lett.}\ }\textbf
  {\bibinfo {volume} {116}},\ \bibinfo {pages} {092402} (\bibinfo {year}
  {2020})}\BibitemShut {NoStop}%
\bibitem [{\citenamefont {Lee}\ \emph {et~al.}(2010)\citenamefont {Lee},
  \citenamefont {Yan}, \citenamefont {Brus}, \citenamefont {Heinz},
  \citenamefont {Hone},\ and\ \citenamefont {Ryu}}]{Lee.2010}%
  \BibitemOpen
  \bibfield  {author} {\bibinfo {author} {\bibfnamefont {Changgu}\ \bibnamefont
  {Lee}}, \bibinfo {author} {\bibfnamefont {Hugen}\ \bibnamefont {Yan}},
  \bibinfo {author} {\bibfnamefont {Louis~E.}\ \bibnamefont {Brus}}, \bibinfo
  {author} {\bibfnamefont {Tony~F.}\ \bibnamefont {Heinz}}, \bibinfo {author}
  {\bibfnamefont {James}\ \bibnamefont {Hone}}, \ and\ \bibinfo {author}
  {\bibfnamefont {Sunmin}\ \bibnamefont {Ryu}},\ }\bibfield  {title} {\enquote
  {\bibinfo {title} {{Anomalous Lattice Vibrations of Single- and Few-Layer
  {MoS$_2$}}},}\ }\href@noop {} {\bibfield  {journal} {\bibinfo  {journal} {ACS
  Nano}\ }\textbf {\bibinfo {volume} {4}},\ \bibinfo {pages} {2695} (\bibinfo
  {year} {2010})}\BibitemShut {NoStop}%
\bibitem [{\citenamefont {Zhu}\ \emph {et~al.}(2020)\citenamefont {Zhu},
  \citenamefont {Li}, \citenamefont {Yang}, \citenamefont {Nie}, \citenamefont
  {Xu}, \citenamefont {Yang}, \citenamefont {Guan}, \citenamefont {Wang},
  \citenamefont {Li}, \citenamefont {Liu}, \citenamefont {Mao}, \citenamefont
  {Xu}, \citenamefont {Yao}, \citenamefont {Yang}, \citenamefont {Shi},
  \citenamefont {Zheng},\ and\ \citenamefont {Jia}}]{Zhu.2020}%
  \BibitemOpen
  \bibfield  {author} {\bibinfo {author} {\bibfnamefont {Zhen}\ \bibnamefont
  {Zhu}}, \bibinfo {author} {\bibfnamefont {Si}~\bibnamefont {Li}}, \bibinfo
  {author} {\bibfnamefont {Meng}\ \bibnamefont {Yang}}, \bibinfo {author}
  {\bibfnamefont {Xiao-Ang}\ \bibnamefont {Nie}}, \bibinfo {author}
  {\bibfnamefont {Hao-Ke}\ \bibnamefont {Xu}}, \bibinfo {author} {\bibfnamefont
  {Xu}~\bibnamefont {Yang}}, \bibinfo {author} {\bibfnamefont {Dan-Dan}\
  \bibnamefont {Guan}}, \bibinfo {author} {\bibfnamefont {Shiyong}\
  \bibnamefont {Wang}}, \bibinfo {author} {\bibfnamefont {Yao-Yi}\ \bibnamefont
  {Li}}, \bibinfo {author} {\bibfnamefont {Canhua}\ \bibnamefont {Liu}},
  \bibinfo {author} {\bibfnamefont {Zhi-Qiang}\ \bibnamefont {Mao}}, \bibinfo
  {author} {\bibfnamefont {Nan}\ \bibnamefont {Xu}}, \bibinfo {author}
  {\bibfnamefont {Yugui}\ \bibnamefont {Yao}}, \bibinfo {author} {\bibfnamefont
  {Shengyuan~A.}\ \bibnamefont {Yang}}, \bibinfo {author} {\bibfnamefont
  {You-Guo}\ \bibnamefont {Shi}}, \bibinfo {author} {\bibfnamefont {Hao}\
  \bibnamefont {Zheng}}, \ and\ \bibinfo {author} {\bibfnamefont {Jin-Feng}\
  \bibnamefont {Jia}},\ }\bibfield  {title} {\enquote {\bibinfo {title} {A
  tunable and unidirectional one-dimensional electronic system
  {Nb$_{2n+1}$Si$_{n}$Te$_{4n+2}$}},}\ }\href@noop {} {\bibfield  {journal}
  {\bibinfo  {journal} {npj Quantum Materials}\ }\textbf {\bibinfo {volume}
  {5}},\ \bibinfo {pages} {35} (\bibinfo {year} {2020})}\BibitemShut {NoStop}%
\bibitem [{\citenamefont {Ohno}(1999)}]{Ohno.1999}%
  \BibitemOpen
  \bibfield  {author} {\bibinfo {author} {\bibfnamefont {Y.}~\bibnamefont
  {Ohno}},\ }\bibfield  {title} {\enquote {\bibinfo {title} {{The
  Scanning-Tunneling Microscopy, the X-Ray Photoelectron Spectroscopy, the
  Inner-Shell-Electron Energy-Loss Spectroscopy Studies of {$M$Te$_2$} and
  {$M$$_3$SiTe$_6$ ($M$= Nb and Ta)}}},}\ }\href@noop {} {\bibfield  {journal}
  {\bibinfo  {journal} {J. Solid State Chem.}\ }\textbf {\bibinfo {volume}
  {142}},\ \bibinfo {pages} {63} (\bibinfo {year} {1999})}\BibitemShut
  {NoStop}%
\bibitem [{\citenamefont {An}\ \emph {et~al.}(2018)\citenamefont {An},
  \citenamefont {Zhang}, \citenamefont {Hu}, \citenamefont {Zhu}, \citenamefont
  {Gao}, \citenamefont {Zhang}, \citenamefont {Xi}, \citenamefont {Ning},
  \citenamefont {Mao},\ and\ \citenamefont {Tian}}]{An.2018}%
  \BibitemOpen
  \bibfield  {author} {\bibinfo {author} {\bibfnamefont {Linlin}\ \bibnamefont
  {An}}, \bibinfo {author} {\bibfnamefont {Hongwei}\ \bibnamefont {Zhang}},
  \bibinfo {author} {\bibfnamefont {Jin}\ \bibnamefont {Hu}}, \bibinfo {author}
  {\bibfnamefont {Xiangde}\ \bibnamefont {Zhu}}, \bibinfo {author}
  {\bibfnamefont {Wenshuai}\ \bibnamefont {Gao}}, \bibinfo {author}
  {\bibfnamefont {Jinglei}\ \bibnamefont {Zhang}}, \bibinfo {author}
  {\bibfnamefont {Chuanying}\ \bibnamefont {Xi}}, \bibinfo {author}
  {\bibfnamefont {Wei}\ \bibnamefont {Ning}}, \bibinfo {author} {\bibfnamefont
  {Zhiqiang}\ \bibnamefont {Mao}}, \ and\ \bibinfo {author} {\bibfnamefont
  {Mingliang}\ \bibnamefont {Tian}},\ }\bibfield  {title} {\enquote {\bibinfo
  {title} {{Magnetoresistance and Shubnikov--de Haas oscillations in layered
  {Na$_3$SiTe$_6$} thin flakes}},}\ }\href@noop {} {\bibfield  {journal}
  {\bibinfo  {journal} {Phys. Rev. B}\ }\textbf {\bibinfo {volume} {97}},\
  \bibinfo {pages} {235133} (\bibinfo {year} {2018})}\BibitemShut {NoStop}%
\bibitem [{\citenamefont {Pang}\ \emph {et~al.}(2020)\citenamefont {Pang},
  \citenamefont {Rezaei}, \citenamefont {Chen}, \citenamefont {Li},
  \citenamefont {Jian}, \citenamefont {Wang}, \citenamefont {Wang},
  \citenamefont {Duan}, \citenamefont {Zebarjadi},\ and\ \citenamefont
  {Yao}}]{Pang.2020}%
  \BibitemOpen
  \bibfield  {author} {\bibinfo {author} {\bibfnamefont {Yahui}\ \bibnamefont
  {Pang}}, \bibinfo {author} {\bibfnamefont {Emad}\ \bibnamefont {Rezaei}},
  \bibinfo {author} {\bibfnamefont {Dongyun}\ \bibnamefont {Chen}}, \bibinfo
  {author} {\bibfnamefont {Si}~\bibnamefont {Li}}, \bibinfo {author}
  {\bibfnamefont {Yu}~\bibnamefont {Jian}}, \bibinfo {author} {\bibfnamefont
  {Qinsheng}\ \bibnamefont {Wang}}, \bibinfo {author} {\bibfnamefont {Zhiwei}\
  \bibnamefont {Wang}}, \bibinfo {author} {\bibfnamefont {Junxi}\ \bibnamefont
  {Duan}}, \bibinfo {author} {\bibfnamefont {Mona}\ \bibnamefont {Zebarjadi}},
  \ and\ \bibinfo {author} {\bibfnamefont {Yugui}\ \bibnamefont {Yao}},\
  }\bibfield  {title} {\enquote {\bibinfo {title} {Thermoelectric properties of
  layered ternary telluride {Na$_3$SiTe$_6$}},}\ }\href@noop {} {\bibfield
  {journal} {\bibinfo  {journal} {Phys. Rev. Materials}\ }\textbf {\bibinfo
  {volume} {4}},\ \bibinfo {pages} {094205} (\bibinfo {year}
  {2020})}\BibitemShut {NoStop}%
\bibitem [{\citenamefont {Ebad-Allah}\ \emph {et~al.}(2019)\citenamefont
  {Ebad-Allah}, \citenamefont {Afonso}, \citenamefont {Krottenm\"uller},
  \citenamefont {Hu}, \citenamefont {Zhu}, \citenamefont {Mao}, \citenamefont
  {Kune\ifmmode~\check{s}\else \v{s}\fi{}},\ and\ \citenamefont
  {Kuntscher}}]{EbadAllah.2019}%
  \BibitemOpen
  \bibfield  {author} {\bibinfo {author} {\bibfnamefont {J.}~\bibnamefont
  {Ebad-Allah}}, \bibinfo {author} {\bibfnamefont {J.~Fern\'andez}\
  \bibnamefont {Afonso}}, \bibinfo {author} {\bibfnamefont {M.}~\bibnamefont
  {Krottenm\"uller}}, \bibinfo {author} {\bibfnamefont {J.}~\bibnamefont {Hu}},
  \bibinfo {author} {\bibfnamefont {Y.~L.}\ \bibnamefont {Zhu}}, \bibinfo
  {author} {\bibfnamefont {Z.~Q.}\ \bibnamefont {Mao}}, \bibinfo {author}
  {\bibfnamefont {J.}~\bibnamefont {Kune\ifmmode~\check{s}\else \v{s}\fi{}}}, \
  and\ \bibinfo {author} {\bibfnamefont {C.~A.}\ \bibnamefont {Kuntscher}},\
  }\bibfield  {title} {\enquote {\bibinfo {title} {Chemical pressure effect on
  the optical conductivity of the nodal-line semimetals {ZrSiY} {(Y=S,Se,Te)}
  and {ZrGeY} {(Y=S,Te)}},}\ }\href@noop {} {\bibfield  {journal} {\bibinfo
  {journal} {Phys. Rev. B}\ }\textbf {\bibinfo {volume} {99}},\ \bibinfo
  {pages} {125154} (\bibinfo {year} {2019})}\BibitemShut {NoStop}%
\bibitem [{\citenamefont {K\"opf}\ \emph {et~al.}(2020)\citenamefont {K\"opf},
  \citenamefont {Ebad-Allah}, \citenamefont {Lee}, \citenamefont {Mao},\ and\
  \citenamefont {Kuntscher}}]{Koepf.2020}%
  \BibitemOpen
  \bibfield  {author} {\bibinfo {author} {\bibfnamefont {M.}~\bibnamefont
  {K\"opf}}, \bibinfo {author} {\bibfnamefont {J.}~\bibnamefont {Ebad-Allah}},
  \bibinfo {author} {\bibfnamefont {S.~H.}\ \bibnamefont {Lee}}, \bibinfo
  {author} {\bibfnamefont {Z.~Q.}\ \bibnamefont {Mao}}, \ and\ \bibinfo
  {author} {\bibfnamefont {C.~A.}\ \bibnamefont {Kuntscher}},\ }\bibfield
  {title} {\enquote {\bibinfo {title} {Influence of magnetic ordering on the
  optical response of the antiferromagnetic topological insulator
  {{MnBi}$_2$Te$_4$}},}\ }\href@noop {} {\bibfield  {journal} {\bibinfo
  {journal} {Phys. Rev. B}\ }\textbf {\bibinfo {volume} {102}},\ \bibinfo
  {pages} {165139} (\bibinfo {year} {2020})}\BibitemShut {NoStop}%
\bibitem [{\citenamefont {Ebad-Allah}\ \emph {et~al.}(2021)\citenamefont
  {Ebad-Allah}, \citenamefont {Rojewski}, \citenamefont {V\"ost}, \citenamefont
  {Eickerling}, \citenamefont {Scherer}, \citenamefont {Uykur}, \citenamefont
  {Sankar}, \citenamefont {Varrassi}, \citenamefont {Franchini}, \citenamefont
  {Ahn}, \citenamefont {Kune\ifmmode~\check{s}\else \v{s}\fi{}},\ and\
  \citenamefont {Kuntscher}}]{EbadAllah.2021}%
  \BibitemOpen
  \bibfield  {author} {\bibinfo {author} {\bibfnamefont {J.}~\bibnamefont
  {Ebad-Allah}}, \bibinfo {author} {\bibfnamefont {S.}~\bibnamefont
  {Rojewski}}, \bibinfo {author} {\bibfnamefont {M.}~\bibnamefont {V\"ost}},
  \bibinfo {author} {\bibfnamefont {G.}~\bibnamefont {Eickerling}}, \bibinfo
  {author} {\bibfnamefont {W.}~\bibnamefont {Scherer}}, \bibinfo {author}
  {\bibfnamefont {E.}~\bibnamefont {Uykur}}, \bibinfo {author} {\bibfnamefont
  {Raman}\ \bibnamefont {Sankar}}, \bibinfo {author} {\bibfnamefont
  {L.}~\bibnamefont {Varrassi}}, \bibinfo {author} {\bibfnamefont
  {C.}~\bibnamefont {Franchini}}, \bibinfo {author} {\bibfnamefont {K.-H.}\
  \bibnamefont {Ahn}}, \bibinfo {author} {\bibfnamefont {J.}~\bibnamefont
  {Kune\ifmmode~\check{s}\else \v{s}\fi{}}}, \ and\ \bibinfo {author}
  {\bibfnamefont {C.~A.}\ \bibnamefont {Kuntscher}},\ }\bibfield  {title}
  {\enquote {\bibinfo {title} {{Pressure-Induced Excitations in the
  Out-of-Plane Optical Response of the Nodal-Line Semimetal {ZrSiS}}},}\
  }\href@noop {} {\bibfield  {journal} {\bibinfo  {journal} {Phys. Rev. Lett.}\
  }\textbf {\bibinfo {volume} {127}},\ \bibinfo {pages} {076402} (\bibinfo
  {year} {2021})}\BibitemShut {NoStop}%
\bibitem [{\citenamefont {Tanner}(2015)}]{Tanner.2015}%
  \BibitemOpen
  \bibfield  {author} {\bibinfo {author} {\bibfnamefont {D.~B.}\ \bibnamefont
  {Tanner}},\ }\bibfield  {title} {\enquote {\bibinfo {title} {{Use of x-ray
  scattering functions in Kramers-Kronig analysis of reflectance}},}\
  }\href@noop {} {\bibfield  {journal} {\bibinfo  {journal} {Phy. Rev. B}\
  }\textbf {\bibinfo {volume} {91}},\ \bibinfo {pages} {035123} (\bibinfo
  {year} {2015})}\BibitemShut {NoStop}%
\bibitem [{\citenamefont {Blaha}\ \emph {et~al.}()\citenamefont {Blaha},
  \citenamefont {Schwarz}, \citenamefont {Madsen}, \citenamefont {Kvasnicka},
  \citenamefont {Luitz}, \citenamefont {Laskowski}, \citenamefont {Tran},\ and\
  \citenamefont {Marks}}]{wien2k}%
  \BibitemOpen
  \bibfield  {author} {\bibinfo {author} {\bibfnamefont {P.}~\bibnamefont
  {Blaha}}, \bibinfo {author} {\bibfnamefont {K.}~\bibnamefont {Schwarz}},
  \bibinfo {author} {\bibfnamefont {G.K.H.}\ \bibnamefont {Madsen}}, \bibinfo
  {author} {\bibfnamefont {D.}~\bibnamefont {Kvasnicka}}, \bibinfo {author}
  {\bibfnamefont {J.}~\bibnamefont {Luitz}}, \bibinfo {author} {\bibfnamefont
  {R.}~\bibnamefont {Laskowski}}, \bibinfo {author} {\bibfnamefont
  {F.}~\bibnamefont {Tran}}, \ and\ \bibinfo {author} {\bibfnamefont {L.D.}\
  \bibnamefont {Marks}},\ }\href@noop {} {}\bibinfo {note} {WIEN2k, An
  Augmented Plane Wave + Local Orbitals Program for Calculating Crystal
  Properties (Karlheinz Schwarz, Techn. Universit\"at Wien, Austria), 2018.
  ISBN 3-9501031-1-2}\BibitemShut {NoStop}%
\bibitem [{\citenamefont {Blaha}\ \emph {et~al.}(2020)\citenamefont {Blaha},
  \citenamefont {Schwarz}, \citenamefont {Tran}, \citenamefont {Laskowski},
  \citenamefont {Madsen},\ and\ \citenamefont {Marks}}]{Blaha2020}%
  \BibitemOpen
  \bibfield  {author} {\bibinfo {author} {\bibfnamefont {P.}~\bibnamefont
  {Blaha}}, \bibinfo {author} {\bibfnamefont {K.}~\bibnamefont {Schwarz}},
  \bibinfo {author} {\bibfnamefont {F.}~\bibnamefont {Tran}}, \bibinfo {author}
  {\bibfnamefont {R.}~\bibnamefont {Laskowski}}, \bibinfo {author}
  {\bibfnamefont {G.~K.~H.}\ \bibnamefont {Madsen}}, \ and\ \bibinfo {author}
  {\bibfnamefont {L.~D.}\ \bibnamefont {Marks}},\ }\bibfield  {title} {\enquote
  {\bibinfo {title} {{WIEN2k: An APW+lo program for calculating the properties
  of solids}},}\ }\href {\doibase 10.1063/1.5143061} {\bibfield  {journal}
  {\bibinfo  {journal} {J. Chem. Phys.}\ }\textbf {\bibinfo {volume} {152}},\
  \bibinfo {pages} {074101} (\bibinfo {year} {2020})}\BibitemShut {NoStop}%
\bibitem [{\citenamefont {Perdew}\ \emph {et~al.}(1996)\citenamefont {Perdew},
  \citenamefont {Burke},\ and\ \citenamefont {Ernzerhof}}]{pbe96}%
  \BibitemOpen
  \bibfield  {author} {\bibinfo {author} {\bibfnamefont {J.~P.}\ \bibnamefont
  {Perdew}}, \bibinfo {author} {\bibfnamefont {K.}~\bibnamefont {Burke}}, \
  and\ \bibinfo {author} {\bibfnamefont {M.}~\bibnamefont {Ernzerhof}},\
  }\bibfield  {title} {\enquote {\bibinfo {title} {Generalized gradient
  approximation made simple},}\ }\href {\doibase 10.1103/PhysRevLett.77.3865}
  {\bibfield  {journal} {\bibinfo  {journal} {Phys. Rev. Lett.}\ }\textbf
  {\bibinfo {volume} {77}},\ \bibinfo {pages} {3865} (\bibinfo {year}
  {1996})}\BibitemShut {NoStop}%
\bibitem [{\citenamefont {Li}\ \emph {et~al.}(1992)\citenamefont {Li},
  \citenamefont {Badding},\ and\ \citenamefont {DiSalve}}]{Li.1992}%
  \BibitemOpen
  \bibfield  {author} {\bibinfo {author} {\bibfnamefont {J.}~\bibnamefont
  {Li}}, \bibinfo {author} {\bibfnamefont {M.~E.}\ \bibnamefont {Badding}}, \
  and\ \bibinfo {author} {\bibfnamefont {F.~J.}\ \bibnamefont {DiSalve}},\
  }\bibfield  {title} {\enquote {\bibinfo {title} {Synthesis and structure of
  {Nb$_3$SiTe$_6$}, a new layered ternary niobium telluride compound},}\
  }\href@noop {} {\bibfield  {journal} {\bibinfo  {journal} {J. Alloys Compd.}\
  }\textbf {\bibinfo {volume} {184}},\ \bibinfo {pages} {257} (\bibinfo {year}
  {1992})}\BibitemShut {NoStop}%
\bibitem [{\citenamefont {Ambrosch-Draxl}\ and\ \citenamefont
  {Sofo}(2006)}]{Draxl2006}%
  \BibitemOpen
  \bibfield  {author} {\bibinfo {author} {\bibfnamefont {C.}~\bibnamefont
  {Ambrosch-Draxl}}\ and\ \bibinfo {author} {\bibfnamefont {J.~O.}\
  \bibnamefont {Sofo}},\ }\bibfield  {title} {\enquote {\bibinfo {title}
  {Linear optical properties of solids within the full-potential linearized
  augmented planewave method},}\ }\href {\doibase 10.1016/j.cpc.2006.03.005}
  {\bibfield  {journal} {\bibinfo  {journal} {Comput. Phys. Commun.}\ }\textbf
  {\bibinfo {volume} {175}},\ \bibinfo {pages} {1--14} (\bibinfo {year}
  {2006})}\BibitemShut {NoStop}%
\bibitem [{Sup()}]{Suppl}%
  \BibitemOpen
  \href@noop {} {}\bibinfo {note} {{See Supplemental Material at --- for
  further information on the experimental data and their
  analysis.}}\BibitemShut {Stop}%
\bibitem [{\citenamefont {Yaresko}\ and\ \citenamefont
  {Pronin}(2021)}]{Yaresko.2021}%
  \BibitemOpen
  \bibfield  {author} {\bibinfo {author} {\bibfnamefont {A.}~\bibnamefont
  {Yaresko}}\ and\ \bibinfo {author} {\bibfnamefont {A.~V.}\ \bibnamefont
  {Pronin}},\ }\bibfield  {title} {\enquote {\bibinfo {title} {{Low-Energy
  Optical Conductivity of {TaP}: Comparison of Theory and Experiment}},}\
  }\href@noop {} {\bibfield  {journal} {\bibinfo  {journal} {Crystals}\
  }\textbf {\bibinfo {volume} {11}},\ \bibinfo {pages} {567} (\bibinfo {year}
  {2021})}\BibitemShut {NoStop}%
\bibitem [{\citenamefont {Crassee}\ \emph {et~al.}(2018)\citenamefont
  {Crassee}, \citenamefont {Martino}, \citenamefont {Homes}, \citenamefont
  {Caha}, \citenamefont {Nov{\'{a} }k}, \citenamefont {Tückmantel},
  \citenamefont {Hakl}, \citenamefont {Nateprov}, \citenamefont {Arushanov},
  \citenamefont {Gibson}, \citenamefont {Cava}, \citenamefont {Koohpayeh},
  \citenamefont {Arpino}, \citenamefont {McQueen}, \citenamefont {Orlita},\
  and\ \citenamefont {Akrap}}]{Crassee.2018}%
  \BibitemOpen
  \bibfield  {author} {\bibinfo {author} {\bibfnamefont {I.}~\bibnamefont
  {Crassee}}, \bibinfo {author} {\bibfnamefont {E.}~\bibnamefont {Martino}},
  \bibinfo {author} {\bibfnamefont {C.~C.}\ \bibnamefont {Homes}}, \bibinfo
  {author} {\bibfnamefont {O.}~\bibnamefont {Caha}}, \bibinfo {author}
  {\bibfnamefont {J.}~\bibnamefont {Nov{\'{a} }k}}, \bibinfo {author}
  {\bibfnamefont {P.}~\bibnamefont {Tückmantel}}, \bibinfo {author}
  {\bibfnamefont {M.}~\bibnamefont {Hakl}}, \bibinfo {author} {\bibfnamefont
  {A.}~\bibnamefont {Nateprov}}, \bibinfo {author} {\bibfnamefont
  {E.}~\bibnamefont {Arushanov}}, \bibinfo {author} {\bibfnamefont {Q.~D.}\
  \bibnamefont {Gibson}}, \bibinfo {author} {\bibfnamefont {R.~J.}\
  \bibnamefont {Cava}}, \bibinfo {author} {\bibfnamefont {S.~M.}\ \bibnamefont
  {Koohpayeh}}, \bibinfo {author} {\bibfnamefont {K.~E.}\ \bibnamefont
  {Arpino}}, \bibinfo {author} {\bibfnamefont {T.~M.}\ \bibnamefont {McQueen}},
  \bibinfo {author} {\bibfnamefont {M.}~\bibnamefont {Orlita}}, \ and\ \bibinfo
  {author} {\bibfnamefont {Ana}\ \bibnamefont {Akrap}},\ }\bibfield  {title}
  {\enquote {\bibinfo {title} {Non-uniform carrier density in {Cd$_3$As$_2$}
  evidenced by optical spectroscopy},}\ }\href {\doibase
  10.1103/physrevb.97.125204} {\bibfield  {journal} {\bibinfo  {journal} {Phys.
  Rev. B}\ }\textbf {\bibinfo {volume} {97}},\ \bibinfo {pages} {125204}
  (\bibinfo {year} {2018})}\BibitemShut {NoStop}%
\bibitem [{\citenamefont {Xu}\ \emph {et~al.}(2020)\citenamefont {Xu},
  \citenamefont {Zhao}, \citenamefont {Yi}, \citenamefont {Wang}, \citenamefont
  {Yin}, \citenamefont {Wang}, \citenamefont {Hu}, \citenamefont {Wang},
  \citenamefont {Liu}, \citenamefont {Xu}, \citenamefont {Lu}, \citenamefont
  {Soluyanov}, \citenamefont {Lei}, \citenamefont {Shi}, \citenamefont {Luo},\
  and\ \citenamefont {Chen}}]{Xu.2020}%
  \BibitemOpen
  \bibfield  {author} {\bibinfo {author} {\bibfnamefont {Y.}~\bibnamefont
  {Xu}}, \bibinfo {author} {\bibfnamefont {J.}~\bibnamefont {Zhao}}, \bibinfo
  {author} {\bibfnamefont {C.}~\bibnamefont {Yi}}, \bibinfo {author}
  {\bibfnamefont {Q.}~\bibnamefont {Wang}}, \bibinfo {author} {\bibfnamefont
  {Q.}~\bibnamefont {Yin}}, \bibinfo {author} {\bibfnamefont {Y.}~\bibnamefont
  {Wang}}, \bibinfo {author} {\bibfnamefont {L.}~\bibnamefont {Hu}}, \bibinfo
  {author} {\bibfnamefont {L.}~\bibnamefont {Wang}}, \bibinfo {author}
  {\bibfnamefont {E.}~\bibnamefont {Liu}}, \bibinfo {author} {\bibfnamefont
  {G.}~\bibnamefont {Xu}}, \bibinfo {author} {\bibfnamefont {L.}~\bibnamefont
  {Lu}}, \bibinfo {author} {\bibfnamefont {A.~A.}\ \bibnamefont {Soluyanov}},
  \bibinfo {author} {\bibfnamefont {H.}~\bibnamefont {Lei}}, \bibinfo {author}
  {\bibfnamefont {Y.}~\bibnamefont {Shi}}, \bibinfo {author} {\bibfnamefont
  {J.}~\bibnamefont {Luo}}, \ and\ \bibinfo {author} {\bibfnamefont {Z.-G.}\
  \bibnamefont {Chen}},\ }\bibfield  {title} {\enquote {\bibinfo {title}
  {{Electronic correlations and flattened band in magnetic Weyl semimetal
  candidate {Co$_3$Sn$_2$S$_2$}}},}\ }\href@noop {} {\bibfield  {journal}
  {\bibinfo  {journal} {Nat. Commun.}\ }\textbf {\bibinfo {volume} {11}},\
  \bibinfo {pages} {3985} (\bibinfo {year} {2020})}\BibitemShut {NoStop}%
\bibitem [{\citenamefont {Shao}\ \emph {et~al.}(2020)\citenamefont {Shao},
  \citenamefont {Rudenko}, \citenamefont {Hu}, \citenamefont {Sun},
  \citenamefont {Zhu}, \citenamefont {Moon}, \citenamefont {Millis},
  \citenamefont {Yuan}, \citenamefont {Lichtenstein}, \citenamefont {Smirnov},
  \citenamefont {Mao}, \citenamefont {Katsnelson},\ and\ \citenamefont
  {Basov}}]{Shao.2020}%
  \BibitemOpen
  \bibfield  {author} {\bibinfo {author} {\bibfnamefont {Y.}~\bibnamefont
  {Shao}}, \bibinfo {author} {\bibfnamefont {A.~N.}\ \bibnamefont {Rudenko}},
  \bibinfo {author} {\bibfnamefont {J.}~\bibnamefont {Hu}}, \bibinfo {author}
  {\bibfnamefont {Z.}~\bibnamefont {Sun}}, \bibinfo {author} {\bibfnamefont
  {Y.}~\bibnamefont {Zhu}}, \bibinfo {author} {\bibfnamefont {S.}~\bibnamefont
  {Moon}}, \bibinfo {author} {\bibfnamefont {A.~J.}\ \bibnamefont {Millis}},
  \bibinfo {author} {\bibfnamefont {S.}~\bibnamefont {Yuan}}, \bibinfo {author}
  {\bibfnamefont {A.~I.}\ \bibnamefont {Lichtenstein}}, \bibinfo {author}
  {\bibfnamefont {D.}~\bibnamefont {Smirnov}}, \bibinfo {author} {\bibfnamefont
  {Z.~Q.}\ \bibnamefont {Mao}}, \bibinfo {author} {\bibfnamefont {M.~I.}\
  \bibnamefont {Katsnelson}}, \ and\ \bibinfo {author} {\bibfnamefont {D.~N.}\
  \bibnamefont {Basov}},\ }\bibfield  {title} {\enquote {\bibinfo {title}
  {{Electronic correlations in nodal-line semimetals}},}\ }\href@noop {}
  {\bibfield  {journal} {\bibinfo  {journal} {Nat. Phys.}\ }\textbf {\bibinfo
  {volume} {16}},\ \bibinfo {pages} {636} (\bibinfo {year} {2020})}\BibitemShut
  {NoStop}%
\bibitem [{\citenamefont {Xu}\ \emph {et~al.}(2018)\citenamefont {Xu},
  \citenamefont {Zhao}, \citenamefont {Marsik}, \citenamefont {Sheveleva},
  \citenamefont {Lyzwa}, \citenamefont {Dai}, \citenamefont {Chen},
  \citenamefont {Qiu},\ and\ \citenamefont {Bernhard}}]{Xu.2018}%
  \BibitemOpen
  \bibfield  {author} {\bibinfo {author} {\bibfnamefont {B.}~\bibnamefont
  {Xu}}, \bibinfo {author} {\bibfnamefont {L.~X.}\ \bibnamefont {Zhao}},
  \bibinfo {author} {\bibfnamefont {P.}~\bibnamefont {Marsik}}, \bibinfo
  {author} {\bibfnamefont {E.}~\bibnamefont {Sheveleva}}, \bibinfo {author}
  {\bibfnamefont {F.}~\bibnamefont {Lyzwa}}, \bibinfo {author} {\bibfnamefont
  {Y.~M.}\ \bibnamefont {Dai}}, \bibinfo {author} {\bibfnamefont {G.~F.}\
  \bibnamefont {Chen}}, \bibinfo {author} {\bibfnamefont {X.~G.}\ \bibnamefont
  {Qiu}}, \ and\ \bibinfo {author} {\bibfnamefont {C.}~\bibnamefont
  {Bernhard}},\ }\bibfield  {title} {\enquote {\bibinfo {title}
  {Temperature-driven topological phase transition and intermediate {Dirac}
  semimetal phase in {ZrTe$_5$}},}\ }\href {\doibase
  10.1103/PhysRevLett.121.187401} {\bibfield  {journal} {\bibinfo  {journal}
  {Phys. Rev. Lett.}\ }\textbf {\bibinfo {volume} {121}},\ \bibinfo {pages}
  {187401} (\bibinfo {year} {2018})}\BibitemShut {NoStop}%
\bibitem [{\citenamefont {Santos-Cottin}\ \emph {et~al.}(2021)\citenamefont
  {Santos-Cottin}, \citenamefont {Casula}, \citenamefont {de' Medici},
  \citenamefont {Le~Mardel\'e}, \citenamefont {Wyzula}, \citenamefont {Orlita},
  \citenamefont {Klein}, \citenamefont {Gauzzi}, \citenamefont {Akrap},\ and\
  \citenamefont {Lobo}}]{Santos-Cottin.2021}%
  \BibitemOpen
  \bibfield  {author} {\bibinfo {author} {\bibfnamefont {David}\ \bibnamefont
  {Santos-Cottin}}, \bibinfo {author} {\bibfnamefont {Michele}\ \bibnamefont
  {Casula}}, \bibinfo {author} {\bibfnamefont {Luca}\ \bibnamefont {de'
  Medici}}, \bibinfo {author} {\bibfnamefont {F.}~\bibnamefont {Le~Mardel\'e}},
  \bibinfo {author} {\bibfnamefont {J.}~\bibnamefont {Wyzula}}, \bibinfo
  {author} {\bibfnamefont {M.}~\bibnamefont {Orlita}}, \bibinfo {author}
  {\bibfnamefont {Yannick}\ \bibnamefont {Klein}}, \bibinfo {author}
  {\bibfnamefont {Andrea}\ \bibnamefont {Gauzzi}}, \bibinfo {author}
  {\bibfnamefont {Ana}\ \bibnamefont {Akrap}}, \ and\ \bibinfo {author}
  {\bibfnamefont {R.~P. S.~M.}\ \bibnamefont {Lobo}},\ }\bibfield  {title}
  {\enquote {\bibinfo {title} {{Optical conductivity signatures of open Dirac
  nodal lines}},}\ }\href@noop {} {\bibfield  {journal} {\bibinfo  {journal}
  {Phys. Rev. B}\ }\textbf {\bibinfo {volume} {104}},\ \bibinfo {pages}
  {{L}201115} (\bibinfo {year} {2021})}\BibitemShut {NoStop}%
\bibitem [{\citenamefont {Ahn}\ \emph {et~al.}(2017)\citenamefont {Ahn},
  \citenamefont {Mele},\ and\ \citenamefont {Min}}]{Ahn.2017}%
  \BibitemOpen
  \bibfield  {author} {\bibinfo {author} {\bibfnamefont {Seongjin}\
  \bibnamefont {Ahn}}, \bibinfo {author} {\bibfnamefont {E.~J.}\ \bibnamefont
  {Mele}}, \ and\ \bibinfo {author} {\bibfnamefont {Hongki}\ \bibnamefont
  {Min}},\ }\bibfield  {title} {\enquote {\bibinfo {title} {Electrodynamics on
  fermi cyclides in nodal line semimetals},}\ }\href {\doibase
  10.1103/PhysRevLett.119.147402} {\bibfield  {journal} {\bibinfo  {journal}
  {Phys. Rev. Lett.}\ }\textbf {\bibinfo {volume} {119}},\ \bibinfo {pages}
  {147402} (\bibinfo {year} {2017})}\BibitemShut {NoStop}%
\bibitem [{\citenamefont {Shao}\ \emph {et~al.}(2019)\citenamefont {Shao},
  \citenamefont {Sun}, \citenamefont {Wang}, \citenamefont {Xu}, \citenamefont
  {Sankar}, \citenamefont {Breidel}, \citenamefont {Cao}, \citenamefont
  {Fogler}, \citenamefont {Millis}, \citenamefont {Chou}, \citenamefont {Li},
  \citenamefont {Timusk}, \citenamefont {Maple},\ and\ \citenamefont
  {Basov}}]{Shao2019}%
  \BibitemOpen
  \bibfield  {author} {\bibinfo {author} {\bibfnamefont {Y.}~\bibnamefont
  {Shao}}, \bibinfo {author} {\bibfnamefont {Z.}~\bibnamefont {Sun}}, \bibinfo
  {author} {\bibfnamefont {Y.}~\bibnamefont {Wang}}, \bibinfo {author}
  {\bibfnamefont {C.}~\bibnamefont {Xu}}, \bibinfo {author} {\bibfnamefont
  {R.}~\bibnamefont {Sankar}}, \bibinfo {author} {\bibfnamefont {A.~J.}\
  \bibnamefont {Breidel}}, \bibinfo {author} {\bibfnamefont {C.}~\bibnamefont
  {Cao}}, \bibinfo {author} {\bibfnamefont {M.~M.}\ \bibnamefont {Fogler}},
  \bibinfo {author} {\bibfnamefont {A.~J.}\ \bibnamefont {Millis}}, \bibinfo
  {author} {\bibfnamefont {F.}~\bibnamefont {Chou}}, \bibinfo {author}
  {\bibfnamefont {Z.}~\bibnamefont {Li}}, \bibinfo {author} {\bibfnamefont
  {T.}~\bibnamefont {Timusk}}, \bibinfo {author} {\bibfnamefont {M.~B.}\
  \bibnamefont {Maple}}, \ and\ \bibinfo {author} {\bibfnamefont {D.~N.}\
  \bibnamefont {Basov}},\ }\bibfield  {title} {\enquote {\bibinfo {title}
  {Optical signatures of dirac nodal lines in {NbAs$_2$}},}\ }\href {\doibase
  10.1073/pnas.1809631115} {\bibfield  {journal} {\bibinfo  {journal} {PNAS}\
  }\textbf {\bibinfo {volume} {116}},\ \bibinfo {pages} {1168} (\bibinfo {year}
  {2019})}\BibitemShut {NoStop}%
\bibitem [{\citenamefont {Carbotte}(2017)}]{Carbotte2017}%
  \BibitemOpen
  \bibfield  {author} {\bibinfo {author} {\bibfnamefont {J.~P.}\ \bibnamefont
  {Carbotte}},\ }\bibfield  {title} {\enquote {\bibinfo {title} {Optical
  response of a line node semimetal},}\ }\href {\doibase
  10.1088/1361-648X/29/4/045301} {\bibfield  {journal} {\bibinfo  {journal} {J.
  Phys. Condens. Matter}\ }\textbf {\bibinfo {volume} {29}},\ \bibinfo {pages}
  {045301} (\bibinfo {year} {2017})}\BibitemShut {NoStop}%
\bibitem [{\citenamefont {Mukherjee}\ and\ \citenamefont
  {Carbotte}(2017)}]{Mukherjee.2017}%
  \BibitemOpen
  \bibfield  {author} {\bibinfo {author} {\bibfnamefont {S.~P.}\ \bibnamefont
  {Mukherjee}}\ and\ \bibinfo {author} {\bibfnamefont {J.~P.}\ \bibnamefont
  {Carbotte}},\ }\bibfield  {title} {\enquote {\bibinfo {title} {Transport and
  optics at the node in a nodal loop semimetal},}\ }\href {\doibase
  10.1103/PhysRevB.95.214203} {\bibfield  {journal} {\bibinfo  {journal} {Phys.
  Rev. B}\ }\textbf {\bibinfo {volume} {95}},\ \bibinfo {pages} {214203}
  (\bibinfo {year} {2017})}\BibitemShut {NoStop}%
\bibitem [{\citenamefont {Martino}\ \emph {et~al.}(2019)\citenamefont
  {Martino}, \citenamefont {Crassee}, \citenamefont {Eguchi}, \citenamefont
  {Santos-Cottin}, \citenamefont {Zhong}, \citenamefont {Gu}, \citenamefont
  {Berger}, \citenamefont {Rukelj}, \citenamefont {Orlita}, \citenamefont
  {Homes},\ and\ \citenamefont {Akrap}}]{Martino.2019}%
  \BibitemOpen
  \bibfield  {author} {\bibinfo {author} {\bibfnamefont {E.}~\bibnamefont
  {Martino}}, \bibinfo {author} {\bibfnamefont {I.}~\bibnamefont {Crassee}},
  \bibinfo {author} {\bibfnamefont {G.}~\bibnamefont {Eguchi}}, \bibinfo
  {author} {\bibfnamefont {D.}~\bibnamefont {Santos-Cottin}}, \bibinfo {author}
  {\bibfnamefont {R.~D.}\ \bibnamefont {Zhong}}, \bibinfo {author}
  {\bibfnamefont {G.~D.}\ \bibnamefont {Gu}}, \bibinfo {author} {\bibfnamefont
  {H.}~\bibnamefont {Berger}}, \bibinfo {author} {\bibfnamefont
  {Z.}~\bibnamefont {Rukelj}}, \bibinfo {author} {\bibfnamefont
  {M.}~\bibnamefont {Orlita}}, \bibinfo {author} {\bibfnamefont {C.~C.}\
  \bibnamefont {Homes}}, \ and\ \bibinfo {author} {\bibfnamefont {Ana}\
  \bibnamefont {Akrap}},\ }\bibfield  {title} {\enquote {\bibinfo {title}
  {Two-dimensional conical dispersion in {ZrTe$_5$} evidenced by optical
  spectroscopy},}\ }\href {\doibase 10.1103/PhysRevLett.122.217402} {\bibfield
  {journal} {\bibinfo  {journal} {Phys. Rev. Lett.}\ }\textbf {\bibinfo
  {volume} {122}},\ \bibinfo {pages} {217402} (\bibinfo {year}
  {2019})}\BibitemShut {NoStop}%
\bibitem [{\citenamefont {Santos-Cottin}\ \emph {et~al.}(2020)\citenamefont
  {Santos-Cottin}, \citenamefont {Martino}, \citenamefont {Le~Mardel\'e},
  \citenamefont {Witteveen}, \citenamefont {von Rohr}, \citenamefont {Homes},
  \citenamefont {Rukelj},\ and\ \citenamefont {Akrap}}]{Santos-Cottin.2020}%
  \BibitemOpen
  \bibfield  {author} {\bibinfo {author} {\bibfnamefont {D.}~\bibnamefont
  {Santos-Cottin}}, \bibinfo {author} {\bibfnamefont {E.}~\bibnamefont
  {Martino}}, \bibinfo {author} {\bibfnamefont {F.}~\bibnamefont
  {Le~Mardel\'e}}, \bibinfo {author} {\bibfnamefont {C.}~\bibnamefont
  {Witteveen}}, \bibinfo {author} {\bibfnamefont {F.~O.}\ \bibnamefont {von
  Rohr}}, \bibinfo {author} {\bibfnamefont {C.~C.}\ \bibnamefont {Homes}},
  \bibinfo {author} {\bibfnamefont {Z.}~\bibnamefont {Rukelj}}, \ and\ \bibinfo
  {author} {\bibfnamefont {Ana}\ \bibnamefont {Akrap}},\ }\bibfield  {title}
  {\enquote {\bibinfo {title} {Low-energy excitations in type-{II} {Weyl}
  semimetal {$T_d$-MoTe$_2$} evidenced through optical conductivity},}\ }\href
  {\doibase 10.1103/PhysRevMaterials.4.021201} {\bibfield  {journal} {\bibinfo
  {journal} {Phys. Rev. Materials}\ }\textbf {\bibinfo {volume} {4}},\ \bibinfo
  {pages} {021201} (\bibinfo {year} {2020})}\BibitemShut {NoStop}%
\bibitem [{\citenamefont {Santos-Cottin}\ \emph {et~al.}(2022)\citenamefont
  {Santos-Cottin}, \citenamefont {Wyzula}, \citenamefont {Le~Mardel\'e},
  \citenamefont {Crassee}, \citenamefont {Martino}, \citenamefont {Nov\'ak},
  \citenamefont {Eguchi}, \citenamefont {Rukelj}, \citenamefont {Novak},
  \citenamefont {Orlita},\ and\ \citenamefont {Akrap}}]{Santos-Cottin.2022}%
  \BibitemOpen
  \bibfield  {author} {\bibinfo {author} {\bibfnamefont {D.}~\bibnamefont
  {Santos-Cottin}}, \bibinfo {author} {\bibfnamefont {J.}~\bibnamefont
  {Wyzula}}, \bibinfo {author} {\bibfnamefont {F.}~\bibnamefont
  {Le~Mardel\'e}}, \bibinfo {author} {\bibfnamefont {I.}~\bibnamefont
  {Crassee}}, \bibinfo {author} {\bibfnamefont {E.}~\bibnamefont {Martino}},
  \bibinfo {author} {\bibfnamefont {J.}~\bibnamefont {Nov\'ak}}, \bibinfo
  {author} {\bibfnamefont {G.}~\bibnamefont {Eguchi}}, \bibinfo {author}
  {\bibfnamefont {Z.}~\bibnamefont {Rukelj}}, \bibinfo {author} {\bibfnamefont
  {M.}~\bibnamefont {Novak}}, \bibinfo {author} {\bibfnamefont
  {M.}~\bibnamefont {Orlita}}, \ and\ \bibinfo {author} {\bibfnamefont {Ana}\
  \bibnamefont {Akrap}},\ }\bibfield  {title} {\enquote {\bibinfo {title}
  {{Addressing shape and extent of Weyl cones in {TaAs} by Landau level
  spectroscopy}},}\ }\href@noop {} {\bibfield  {journal} {\bibinfo  {journal}
  {Phys. Rev. B}\ }\textbf {\bibinfo {volume} {105}},\ \bibinfo {pages}
  {{L}081114} (\bibinfo {year} {2022})}\BibitemShut {NoStop}%
\bibitem [{\citenamefont {Uykur}\ \emph {et~al.}(2021)\citenamefont {Uykur},
  \citenamefont {Ortiz}, \citenamefont {Iakutkina}, \citenamefont {Wenzel},
  \citenamefont {Wilson}, \citenamefont {Dressel},\ and\ \citenamefont
  {Tsirlin}}]{Uykur.2021}%
  \BibitemOpen
  \bibfield  {author} {\bibinfo {author} {\bibfnamefont {E.}~\bibnamefont
  {Uykur}}, \bibinfo {author} {\bibfnamefont {B.~R.}\ \bibnamefont {Ortiz}},
  \bibinfo {author} {\bibfnamefont {O.}~\bibnamefont {Iakutkina}}, \bibinfo
  {author} {\bibfnamefont {M.}~\bibnamefont {Wenzel}}, \bibinfo {author}
  {\bibfnamefont {S.~D.}\ \bibnamefont {Wilson}}, \bibinfo {author}
  {\bibfnamefont {M.}~\bibnamefont {Dressel}}, \ and\ \bibinfo {author}
  {\bibfnamefont {A.~A.}\ \bibnamefont {Tsirlin}},\ }\bibfield  {title}
  {\enquote {\bibinfo {title} {{Low-energy optical properties of the
  nonmagnetic kagome metal {CsV$_3$Sb$_5$}}},}\ }\href@noop {} {\bibfield
  {journal} {\bibinfo  {journal} {Phys. Rev. B}\ }\textbf {\bibinfo {volume}
  {104}},\ \bibinfo {pages} {045130} (\bibinfo {year} {2021})}\BibitemShut
  {NoStop}%
\bibitem [{\citenamefont {Uykur}\ \emph {et~al.}(2022)\citenamefont {Uykur},
  \citenamefont {Ortiz}, \citenamefont {Wilson}, \citenamefont {Dressel},\ and\
  \citenamefont {Tsirlin}}]{Uykur2022}%
  \BibitemOpen
  \bibfield  {author} {\bibinfo {author} {\bibfnamefont {Ece}\ \bibnamefont
  {Uykur}}, \bibinfo {author} {\bibfnamefont {Brenden~R.}\ \bibnamefont
  {Ortiz}}, \bibinfo {author} {\bibfnamefont {Stephen~D.}\ \bibnamefont
  {Wilson}}, \bibinfo {author} {\bibfnamefont {Martin}\ \bibnamefont
  {Dressel}}, \ and\ \bibinfo {author} {\bibfnamefont {Alexander~A.}\
  \bibnamefont {Tsirlin}},\ }\bibfield  {title} {\enquote {\bibinfo {title}
  {Optical detection of the density-wave instability in the kagome metal
  {KV$_3$Sb$_5$}},}\ }\href {\doibase 10.1038/s41535-021-00420-8} {\bibfield
  {journal} {\bibinfo  {journal} {npj Quantum Materials}\ }\textbf {\bibinfo
  {volume} {7}},\ \bibinfo {pages} {16} (\bibinfo {year} {2022})}\BibitemShut
  {NoStop}%
\bibitem [{\citenamefont {Wenzel}\ \emph {et~al.}(2022)\citenamefont {Wenzel},
  \citenamefont {Ortiz}, \citenamefont {Wilson}, \citenamefont {Dressel},
  \citenamefont {Tsirlin},\ and\ \citenamefont {Uykur}}]{Wenzel2022}%
  \BibitemOpen
  \bibfield  {author} {\bibinfo {author} {\bibfnamefont {M.}~\bibnamefont
  {Wenzel}}, \bibinfo {author} {\bibfnamefont {B.~R.}\ \bibnamefont {Ortiz}},
  \bibinfo {author} {\bibfnamefont {S.~D.}\ \bibnamefont {Wilson}}, \bibinfo
  {author} {\bibfnamefont {M.}~\bibnamefont {Dressel}}, \bibinfo {author}
  {\bibfnamefont {A.~A.}\ \bibnamefont {Tsirlin}}, \ and\ \bibinfo {author}
  {\bibfnamefont {E.}~\bibnamefont {Uykur}},\ }\bibfield  {title} {\enquote
  {\bibinfo {title} {Optical study of {RbV$_3$Sb$_5$}: Multiple density-wave
  gaps and phonon anomalies},}\ }\href {\doibase 10.1103/PhysRevB.105.245123}
  {\bibfield  {journal} {\bibinfo  {journal} {Phys. Rev. B}\ }\textbf {\bibinfo
  {volume} {105}},\ \bibinfo {pages} {245123} (\bibinfo {year}
  {2022})}\BibitemShut {NoStop}%
\end{thebibliography}
\end{document}